\documentstyle[psfig]{mn}

\input{epsf}

\def\lsim{\mathrel{\hbox{\rlap{\hbox{\lower4pt\hbox{$\sim$}}}\hbox{$<$}}}}
\def\gsim{\mathrel{\hbox{\rlap{\hbox{\lower4pt\hbox{$\sim$}}}\hbox{$>$}}}}

\def\and {\rm {et al.} \rm} 

\begin{document}

\title[Spatial and Dynamical properties of Voids] 
{Spatial and Dynamical Properties of Voids in a $\Lambda$CDM Universe}
\author[Padilla et al.]{
\parbox[t]{\textwidth}{
Padilla, N.D.$^{1}$, Ceccarelli, L.$^2$, \& Lambas, D.G.$^2$}
\vspace*{6pt} \\ 
$^1$Departamento de Astronom\'\i a y Astrof\'\i sica, Pontificia
Universidad Cat\'olica, V. Mackenna 4860, Santiago 22, Chile.\\
$^2$IATE, Observatorio Astron\'omico de la Universidad Nacional de C\'ordoba,
    Laprida 851, 5000, C\'ordoba, Argentina. \\ 
}

\maketitle

\begin{abstract}
We study statistical properties of voids
in the distribution
of mass, dark-matter haloes and galaxies ($B_J<-16$) in 
a $\Lambda$CDM numerical simulation populated with galaxies
using a semi-analytic galaxy formation model(GALFORM, Cole et al. 2000). 
We find that the properties 
of voids selected from GALFORM galaxies are compatible with those of voids
identified from a population of haloes with mass $M>10^{11.5}$h$^{-1}$M$_{\odot}$,
similar to the median halo mass, $M_{\rm med}=10^{11.3}$h$^{-1}$M$_{\odot}$.  We also find that the number
density of galaxy- and halo-defined voids can be up to two orders of magnitude higher than
mass-defined voids for large void radii, however, 
we observe that this difference is reduced to about half an order
of magnitude when the positions are considered in
redshift-space.  As expected, there are outflow velocities which
show their maximum at larger void-centric distances for larger voids. 
We find a linear relation for the maximum outflow velocity,
$v_{\rm max}=v_0 r_{\rm void}$.  The 
void-centric distance where this maximum occurs,
follows a suitable power law fit of the
form, $\log(d_{v \rm max})=(r_{\rm void}/A)^B$.
At sufficiently large distances, we find mild infall motions onto the
sub-dense regions.
The galaxy velocity field around galaxy-defined voids is consistent with the results
of haloes with masses above the median, 
showing milder outflows than the mass around mass-defined voids.  
We find that a similar analysis in redshift 
space would make both outflows and infalls to appear with a lower amplitude.  
We also find that the velocity dispersion of galaxies and haloes
is larger in the direction parallel to the void walls by
$\simeq10-20\%$.
Given that voids are by definition sub-dense regions,
the cross-correlation function between galaxy-defined voids and galaxies 
are close to $\xi=-1$ out to separations comparable 
to the void size, and at
larger separations the correlation function level increases 
approaching the values of the auto-correlation function of galaxies.
The cross-correlation amplitude of mass-defined voids vs. mass has a more gentle
behaviour remaining negative at larger distances.  
The cross- to auto-correlation function ratio as a function of
the distance normalised to the void radius,
shows a small scatter around a relation that depends only on the object
used to define the voids (galaxies or haloes for instance).
The distortion pattern observed in
$\xi(\sigma,\pi)$ is that of an elongation along the line of sight
which extends out to large separations.  Positive $\xi$ contours
evidence finger-of-god motions at the void walls.  Elongations along
the line of sight are roughly comparable between galaxy-, halo- and 
mass-defined voids.
\end{abstract}

\begin{keywords}
large-scale structure of Universe, methods: N-body simulations,
galaxies: kinematics and dynamics
cosmology: theory
\end{keywords}

\section{Introduction}
\label{sec:intro}

Voids can be thought of as large volumes with very low galaxy density
surrounded by the walls and filaments of the cosmic web.  These objects were
first detected by Gregory \& Thompson (1978), 
Joeever, Einasto \& Tago (1978), Kirshner et al. (1981)
and Geller \& Huchra (1989).  Soon after their discovery,
White (1979), Hoffman \& Shaham (1982) and Peebles (1982)
derived theoretical statistical properties of voids and
related them to the gravitational instability scenario.
Einasto, Einasto \& Gramann (1989) also estimated sizes 
of voids in different samples of galaxies in
redshift surveys and compared their results with numerical
simulations.
The characteristics of voids in the galaxy distribution, such as
the void probability distribution were studied by
Vogeley et al. (1994), Ghigna et al. (1996), and M\"uller et al. (2000),
whereas the fraction of the total volume in the Universe occupied by 
voids, which was found to add up to about $50\%$, was measured by
El-Ad \& Piran (1997, 2000),
Plionis \& Basilakos (2002), and Hoyle \& Vogeley (2002).
Peebles (2001) presents a detailed
review of the observational works devoted to void statistics.

More recently, theoretical studies on the characteristics 
of voids and their surrounding
structure have concentrated on several different aspects.  For instance,
van de Weygaert (2004) presented an excursion set approach (Bond et al. 1991)
for predicting the distribution function of void sizes and its evolution
with redshift.  Patiri et al. (2004) also carry out an excursion set
study on voids, and are able to characterize the number density of voids as a function
of void radius as well as the mass function of haloes in voids.  Their
model takes into account the fact that the properties of voids 
depend on the haloes used to identify them (Gottl\"ober et al., 2003, and
references therein).  Based on this result, Patiri et al. 
provide a framework which allows a comparison between theoretical
predictions and statistical measures of voids found in the distribution
of the visible galaxies that populate haloes.  
Also, taking into account the nearly spherical average 
shape of voids,  Ryden (1995) showed that it is possible to 
use the cosmologically induced redshift-space distortion in the apparent shape of
intermediate redshift voids to measure the deceleration parameter, $q_0$.

On the observational side, only very recently it has become possible
to perform accurate statistical studies of voids using real data.  
For instance, Croton et al. (2004) used the 2-degree 
Field Galaxy Redshift Survey (2dFGRS) to measure the reduced void probability
function (RVPF), which connects the distribution of voids to the
higher moments of galaxy counts in cells.  
These new redshift surveys also allow the study of the population of galaxies
inside voids.  Hoyle et al. (2003) use the Sloan Digitized Sky Survey (SDSS) 
to study the luminosity function of
void galaxies, which they find by setting an upper
limit on the local density contrast at the galaxy position,
$\delta \rho/\rho\leq-0.6$.
They find that galaxies are fainter in voids, but that the
faint-end slope of the luminosity function, is consistent with that of
galaxies in higher density environments.  Also using the
SDSS, Goldberg et al. (2004) study the mass function of void galaxies
selected using the same density threshold as Hoyle et al. (2003),
finding that galaxies in voids are nearly unbiased with respect to
the mass.  

In this paper we carry out a comprehensive numerical study of the
dynamical and spatial properties of galaxies, dark-matter particles
and haloes with respect to void centres.  
We study how the properties
of voids identified from the distribution of dark-matter particles
and haloes in a simulation
relate to those of voids identified from the galaxy distribution.  
Goldberg et al. find that galaxies trace
the distribution of mass within voids, a fact that contrasts
with the large biases expected in higher density environments, 
such as walls, filaments and clusters.  Since these objects 
set the boundaries of voids, this study provides a useful insight
on the global segregation pattern of dark-matter and galaxies.  
The study of the effects of observational biases occurring in
flux limited redshift surveys, and an application of the statistics
presented here to observational catalogues are carried out in 
a forthcoming paper, Ceccarelli et al. (2004a).

This paper is organised as follows.  Section \ref{sec:voids}
describes the semi-analytical numerical simulation used in this work
and the procedure adopted for identifying voids in the simulation box.
Section \ref{sec:vpecs} analyses the outflow signatures that characterise
the regions surrounding voids.  In section \ref{sec:correlations} we
study the spatial 2-point cross-correlation functions between voids
and different tracers, dark-matter particles, haloes and galaxies.  
Finally, section \ref{sec:conc} summarises
our results and presents the main conclusions extracted from this work.

\section{Voids in the numerical simulation}
\label{sec:voids}

In this work we study void statistics using a $\Lambda$CDM
numerical simulation which contains $125$ million dark matter particles and 
$\sim 2$ million galaxies from the GALFORM semi-analytic galaxy formation model 
(Cole et al. 2000), kindly provided by the Durham group.
There are approximately
$650,000$ dark-matter haloes with at least $10$ members,
identified from the distribution of datk-matter particles in the simulation
using a Friends-of-Friends (FOF from now on)
algorithm with a linking length $b_{\rm FOF}=0.2$.  
The resulting haloes are characterised by a minimum
mass of $M_{\rm min}=1.05\ 10^{11}$h$^{-1}$M$_{\odot}$,
a median mass $M_{\rm med}=2\ 10^{11}$h$^{-1}$M$_{\odot}$,
and a maximum $M_{\rm max}=2\ 10^{15}$h$^{-1}$M$_{\odot}$.
It should
be noticed that the FOF algorithm does not resolve substructure
within the dark-matter haloes.
The semi-analytic model output consists of approximately $2$ million galaxies 
with $B_J<-16$ at $z=0$ 
(from now on, we will refer to all galaxies above this magnitude
limit, simply as galaxies) in the simulation.  The $25$, $50$ and $75$
percentiles of the haloes hosting at least one galaxy are 
$1.57\ 10^{11}$h$^{-1}$M$_{\odot}$,
$3.17\ 10^{11}$h$^{-1}$M$_{\odot}$, and
$8.95\ 10^{11}$h$^{-1}$M$_{\odot}$ respectively.
We now describe the simulation
parameters: the box side measures $250$h$^{-1}$Mpc, the
matter density parameter corresponds to $\Omega_m=0.3$,
the value of the vacuum density parameter is $\Omega_{\Lambda}=0.7$, 
the Hubble constant, $H=h 100$kms$^{-1}$Mpc$^{-1}$, with $h=0.7$,
and the primordial power spectrum slope is $n_s=0.97$.  The
present day amplitude of fluctuations in spheres of
$8$h$^{-1}$Mpc is set to $\sigma_8=0.8$.  This particular
cosmology is in line with recent cosmic microwave background
anisotropy (WMAP, Spergel et al. 2003) and large scale structure 
(2dFGRS, Percival et al. 2004) measurements.  

We use this simulation to study the dynamical and spatial characteristics
of galaxies, dark-matter haloes and dark-matter particles around voids.  Throughout 
this paper, errors in the statistics are calculated using the jackknife method, 
which has been shown to provide equivalent errors to those found by applying 
the statistics to a large number of independent simulations and measuring
a variance (see for instance, Croton et al. 2004, Padilla \& Baugh, 2003).  
These results indicate that jackknife errors provide a reasonable estimate of
statistical uncertainties and cosmic variance.  

The use of semi-analytic numerical simulations
allows a comparison of galaxy-, halo- and mass-defined voids, since the population of
galaxies in the simulation box is physically motivated, 
subject to evolution via mergers, metal enrichment, dust evolution,
and other important astrophysical processes (for full details
see Cole et al. 2000). 

Throughout this paper we compare results from real- and redshift-space derived
quantities.  When referring to voids this indicates that,
when identifying voids in the simulation, the positions
of particles, haloes or galaxies were considered either in real- or redshift-space.
We implement redshift-space
positioning in this simulation by displacing the $z-$coordinates of 
particles and galaxies by the $z-$component of their velocities in units of 
h$^{-1}$Mpc (distant observer looking down the $z-$axis).

In the next subsections
we describe the void finding algorithm, present the resulting
number density of voids as a function of void radius, and calculate the 
void auto-correlation function.

\begin{figure}
{\epsfxsize=8.truecm 
\epsfbox[20 20 575 575]{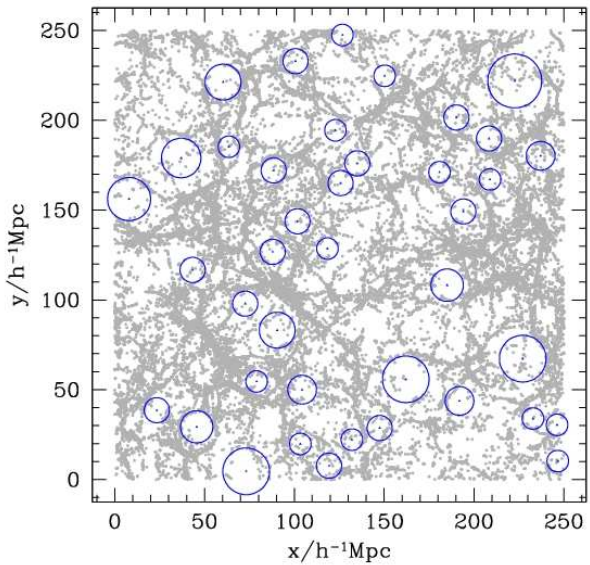}}
\caption{
Slice of the numerical simulation box, corresponding
to $100<z/$h$^{-1}$Mpc$<110$, showing the
positions of semi-analytic galaxies (black dots) and voids
found from the galaxy positions
(gray dots).  
The circles indicate the spatial extent of the
voids.
}
\label{fig:pos}
\end{figure}

\begin{figure}
{\epsfxsize=11.truecm 
\epsfbox[20 160 575 705]{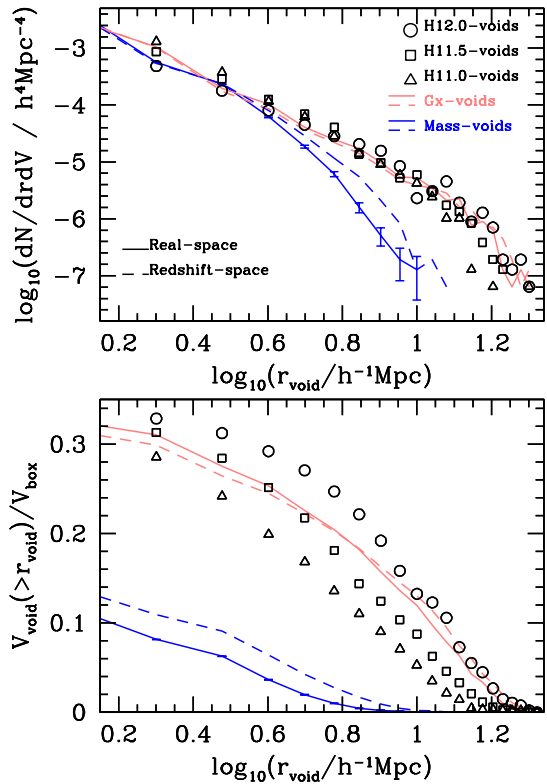}}
\caption{
Upper panel: Number density of voids as a function of void size.
The blue (black) lines correspond to voids found from the distribution
of dark-matter in the simulation box.  Red (gray) lines show the
results obtained from the simulation galaxies.   The symbols
represent results from the distribution of H$11$-, H$11.5$- and H$12$-defined voids
(triangles, squares and circles respectively).  The solid and dashed lines
correspond to the number density of voids found from positions of objects
in real- and redshift-space respectively.
Lower Panel: cumulative volume fraction in voids of radius $r>r_{\rm void}$.
Lines, colours and symbols are as in the upper panel.
}
\label{fig:numdens}
\end{figure}

\subsection{Void finding algorithm}

We find voids in the simulation by identifying spherical volumes where
the overall density contrast satisfies $\delta=-0.9$, 
which corresponds to a usual definition of an under-dense region.  Later
in this section we find that small variations on this condition do not
affect significantly our statistics.
We use the mass, dark-matter haloes, and the galaxies in the simulation
to find voids, and we also investigate in a simple way the
effects of redshift space distortions in the identification of voids.
Therefore, we produce ten different catalogues of voids in the
numerical simulation.  We first have $5$ sets of voids,
found in the distribution of dark-matter 
(which we will refer to as mass-defined voids), haloes with masses
above $10^{11}$, $10^{11.5}$ and $10^{12}$h$^{-1}$M$_{\sun}$ (halo-defined voids, also
referred to as H$11$-, H$11.5$- and H$12$-defined voids respectively)\footnote{
The halo mass thresholds are selected so as to have samples of haloes with
space densities differing by half an order of magnitude; 
H11 haloes characterised by $n=5 \ 10^{-3}$, H11.5 by $n=1.5\ 10^{-3}$, and H12 haloes 
by $n=5\ 10^{-4}$}, and voids found in the distribution of galaxies
(referred to as galaxy-defined voids) in real space.  Repeating the identification in
redshift-space provides us with the remaining $5$ samples.  

For each of these different samples of voids, the procedure
followed is exactly the same, and comprises the following steps:
\begin{itemize}
\item We produce a large number of random positions which will be used
to test whether the density contrast in spheres centred on these positions
satisfies the condition $\delta<-0.9$.  The higher the
number of random positions the higher the accuracy in finding the true
centre of a void in the simulation.  
\item We calculate the density contrast as a function of radius
of a sphere centred on the random positions generated in the first step.
\item We identify the radii for which the void density criterion $\delta<-0.9$ is
satisfied.  In the event where this condition is satisfied more than once
at different radii, we only consider the largest radius.
\item Up to this point, the outcome of the identification process will likely consist of
a large number of underdense overlaping spheres.
Therefore, the last step consists of removing all spheres overlapping with a
larger void.  For instance, this last step will make a large underdense 
spherical region of radius $r$ be represented by a single sphere of radius 
$r$, instead of by many spheres of smaller radius that could be
able to fill the underdense volume. 
\end{itemize}

The nature of the finding algorithm makes
voids of comparable sizes at similar positions be simply replaced by
a larger void which naturally occupies the volume of all the 
smaller voids together.

We have made several checks to make sure the void finding algorithm
is reliable.  We have varied the upper limit in $\delta$ up to
$\delta<-0.6$, and find that in general results are not significantly
affected in the case of galaxy- and halo-defined voids.  The main difference resides in
the appearance of large spurious ($r_{\rm void}>30$h$^{-1}$Mpc) voids.

We have also checked that the removal of overlapping voids does
not affect significantly the mean values of the statistics we investigate
in this work by allowing for instance, voids within voids.  It is important, 
however, to take this effect into
account when calculating errors, since many slightly de-centred voids representing
only one physical void in the simulation artificially lowers the size
of statistical errors.

Given that dark-matter particles outnumber galaxies by a factor
$\simeq 100$ in the simulation, we tested for the impact of the number of 
particles in the identification procedure.  In order to do this we
select $1/100$th of the total dark-matter particles and run our void 
identification algorithm.  The resulting void catalogue is almost
identical to the one found from the full simulation.

Figure \ref{fig:pos} shows a slice of the simulation box
with the distribution of semi-analytic galaxies and
the positions of voids found by our procedure.  In order to
improve clarity, in this particular
plot we only show voids with radii in the range $10<r_{\rm void}
/$h$^{-1}$Mpc.  As can be seen, the void identification algorithm
is quite satisfactory, picking out low density regions with void
radii compatible with the observed galaxy distribution.  The
empty spaces not filled with voids are expected for instance
from voids with $r_{\rm void}<10$h$^{-1}$Mpc, or voids 
with centres outside the slice shown in this figure.

\subsection{Void number density}

The first statistic we measure using the voids identified in 
the previous subsection is the number density as a function
of void radius, $r_{\rm void}$.  We show the measured results from the 
simulation box in the upper panel of figure \ref{fig:numdens},
where the red (gray) lines correspond to voids found from the spatial
distribution of galaxies and the blue (black) lines from the mass.
Solid lines
correspond to real-space, dashed lines to redshift-space.
Symbols show the results from the distribution of voids identified
from haloes of different masses as indicated in the key. 
It is noticeable the good agreement between these estimates
for $r<4$h$^{-1}$Mpc, where all the void number densities are virtually
indistinguishable.  At larger object/void-centre separations
the different estimates start to diverge.  For instance
the difference between real- and redshift-space estimates can
diverge by as much as a factor of $\simeq 2.5$, as is the case
at large mass-void radii $r_{\rm void}>10$h$^{-1}$Mpc. This 
is attributable to the fingers
of god effect which takes place in the void walls and displaces 
dark-matter particles from the walls and dilutes the
void boundaries.  Galaxy- and halo-void number densities
are not significantly affected by a redshift-space identification
of voids as expected, since the peculiar velocities of mass and
galaxies are small ($v/H_0~1h^{-1}Mpc$) compared to the void
sizes. 
The difference between number densities of voids
found from the distribution of galaxies and mass starts to be noticeable
at separations $> 4$h$^{-1}$Mpc; eventually there are no longer
mass-defined voids with $r_{\rm void}>10$h$^{-1}$Mpc whereas it is still
possible to identify galaxy-defined voids with $r_{\rm void}>20$h$^{-1}$Mpc.
The number density of halo-defined voids is more comparable to that
of galaxy-defined voids.

The increase in the number density of voids with the mass of
the haloes used to identify them can be understood in terms of
the bias factor between haloes and the mass.  Higher mass
haloes are more strongly biased with respect to the mass,
and therefore are preferably found in high mass density
regions, which correspond preferably to the centres of void walls and filaments.  
The net effect this would have is to increase the radius
of the void when identified from a biased population of objects,
and would explain the apparent increase in number density
observed in figure \ref{fig:numdens}.

From the number density of voids, it is straight-forward to
obtain the fraction of volume inside voids in the simulation.
The lower panel of fig. \ref{fig:numdens},
shows the volume inside voids with $r>r_{\rm void}$.
As can be seen, the difference between
mass- and galaxy-defined voids is even more evident here.  Also,
it is now possible to recognise a good agreement between
galaxy-defined voids and H$11.5$- for $r_{void}<6$h$^{-1}$Mpc and 
H$12$-defined with $r_{void}>6$h$^{-1}$Mpc voids. 

This demonstrates that caution should be taken when analysing properties
of galaxy-defined voids, which can differ from those of halo- and mass-defined voids.
As mentioned above,
the reason for this discrepancy is the slightly biased nature of the
galaxy distribution in this simulation with respect to the dark-matter 
particles and haloes.  
In particular, the spatial distribution of galaxies shows larger and
more numerous voids than
the distributions of mass or low mass haloes 
(less biased populations).
Given that the redshift-space effect enhances the number density of mass-defined voids
more strongly, the differences between results
from mass- and galaxy-defined voids are less important in redshift-space.

The volume
fraction inside voids can be as large as $30\%$, a value
which is slightly lower than estimates from redshift
surveys of $50\%$ by El-Ad \& Piran, 1997, 2000,
Plionis \& Basilakos, 2002, and Hoyle \& Vogeley, 2002.
As pointed out above, our estimate of the volume fraction
is only an order of magnitude estimate at best, since our simulated
volume is limited and we might be badly affected by sampling
statistics in the large-volume end of the void population.

\begin{figure}
\begin{picture}(230,440)
\put(0,0){\psfig{file=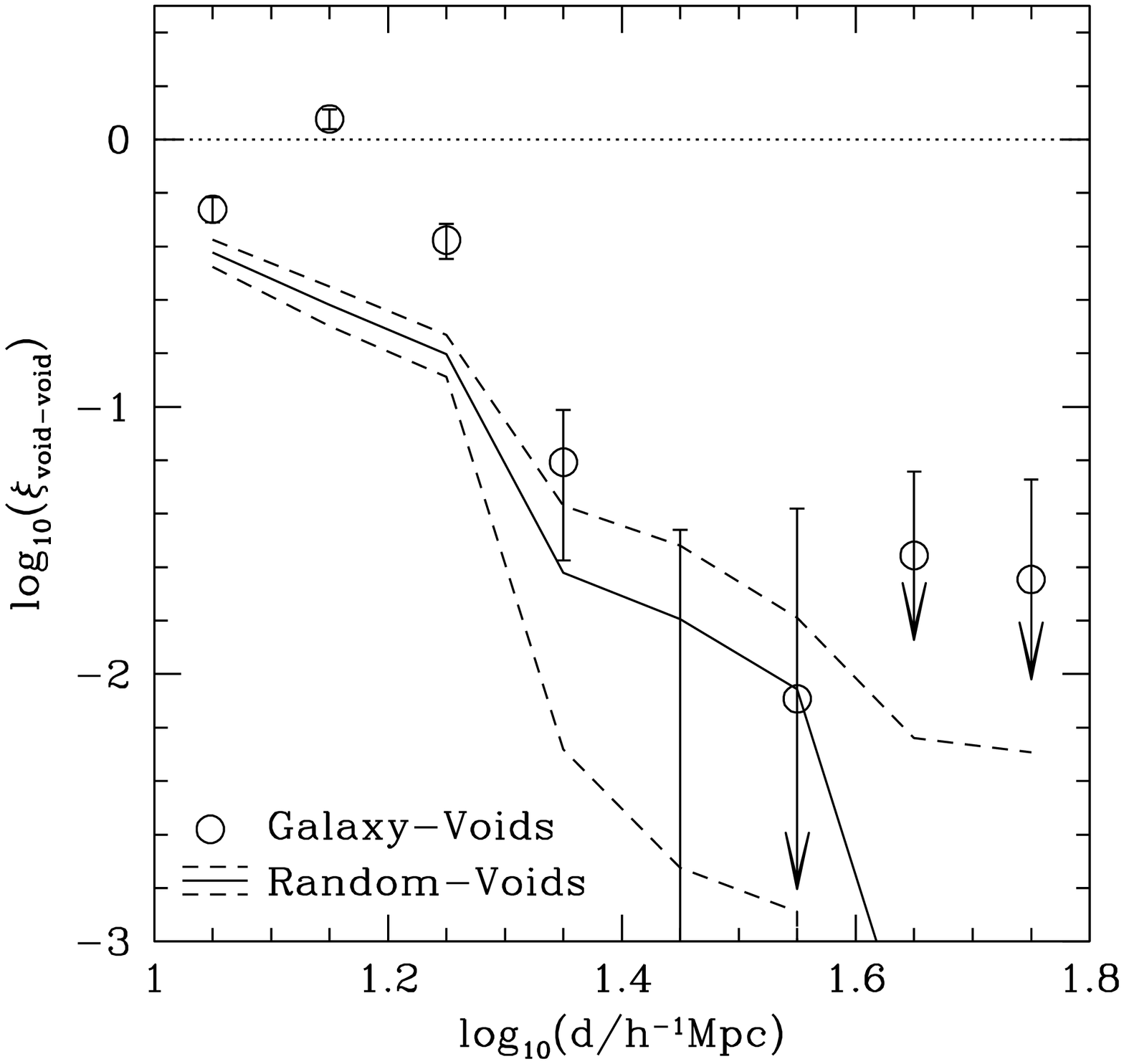,width=8.cm}}
\put(-7,165){\psfig{file=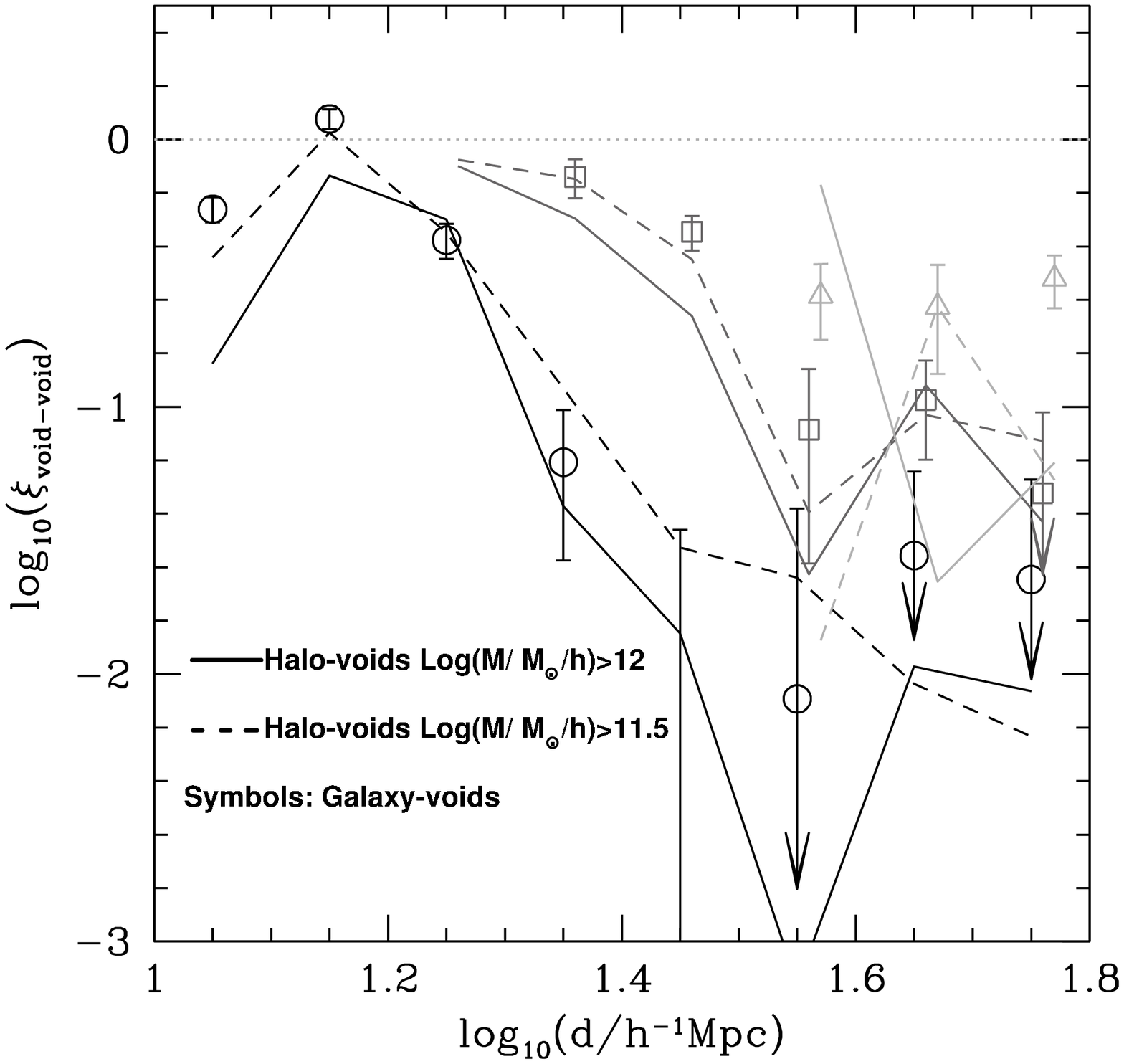,width=8.cm}}
\end{picture}
\caption{
Top panel:
void 2-point auto-correlation function for voids with
radii in the ranges $4<r_{\rm void}/h^{-1}Mpc<8$ (Black lines and
circles), $8<r_{\rm void}/h^{-1}Mpc<12$ (dark-grey lines and squares), 
and $16<r_{\rm void}/h^{-1}Mpc<20$ (light-grey lines and triangles).
The symbols represent results from galaxy-defined voids, the
lines from halo-defined voids.
Bottom panel: comparison between the auto-correlation
function of galaxy- and random-voids
with the same number density as a function of void radius.
}
\label{fig:xivoids}
\end{figure}

\subsection{Void 2-point auto-correlation function}

A simple statistics that can be applied to the void positions is the
2-point auto-correlation function. The advantages in studying the
correlation function relies in its widespread use and in its direct
connection to the power spectrum of density fluctuations.  

The upper panel of figure \ref{fig:xivoids} shows the auto-correlation function
of galaxy- (symbols) and H$12$- and H$11.5$-defined voids (dashed, $M>10^{12}$h$^{-1}$M$_{\sun}$,
solid, $M>10^{11.5}$h$^{-1}$M$_{\sun}$) for different
void radii (gray scale and different symbols as indicated in the figure caption)
\footnote{
The number of centres used in the computation of the void-autocorrelation
function is roughly $3000$, $280$, and $15$, for the three ranges of
void radii shown in figure \ref{fig:xivoids} (smaller to larger void radii, respectively),
which explains the noise present in the correlation functions of voids
of large radius.
}.  
As can be seen,  larger voids are
more strongly clustered, and show a positive correlation signal out to
larger separations.  The amplitude of these correlations, however,
is quite low and we only find a significant signal out to scales
$\simeq 50$h$^{-1}$Mpc.  It is interesting to note that 
at scales larger than $~15$h$^{-1}$Mpc, the slopes
of the correlation functions for voids of different radii are
similar.
The auto-correlation function of voids defined using the most
massive haloes (H$12$, dashed lines) provides the best match to the
clustering of galaxy-defined voids.

In order to assess whether the correlation function responds to
an effect of the void identification procedure, we
show in the lower panel of figure \ref{fig:xivoids}
the average correlation function (and $1-\sigma$ confidence
region) of $100$ sets of randomly placed spheres (random-voids).  Each
of these sets has the same number density as a function of radius
than galaxy-defined voids, and also follows the exclusion
constraints set for the identification of galaxy-defined voids.  Also,
random-voids are placed in a volume with the same size as the 
simulation box.  As can be seen, the exclusion constraint alone 
can produce a non-zero correlation function consistent with a
power law of index $\simeq-3$.  The
comparison with the galaxy-defined voids correlation function, however,
shows that the amplitude of the galaxy-void correlation function 
is still significantly higher than that of the random-voids.  This 
could be interpreted as evidence of the influence of hierarchical 
clustering in the spatial distribution of galaxy-defined voids.

We also calculated the correlation function of sets of random-voids
with a different number density as a function of void radius,
in particular, similar to that of H$11$-defined voids.  We find that
the comparison between H$11$-defined voids and this new set of 
random-voids is identical to the comparison we carry out above,
namely that the correlation function of H$11$-defined voids is 
higher in amplitude than that of random-voids for small separations.

\vskip .5cm

Our preliminary conclusion from the study of the number
density of voids and void auto-correlation functions 
in the numerical simulation is that
galaxy-defined voids show good agreement with voids identified from moderate to high
mass haloes ($M>10^{11.5}$h$^{-1}$M$_{\sun}$).  From now on,
we will mostly compare results
from these two sets of voids, unless we state otherwise.

\begin{figure*}
\begin{picture}(430,250)
\put(0,-20){\psfig{file=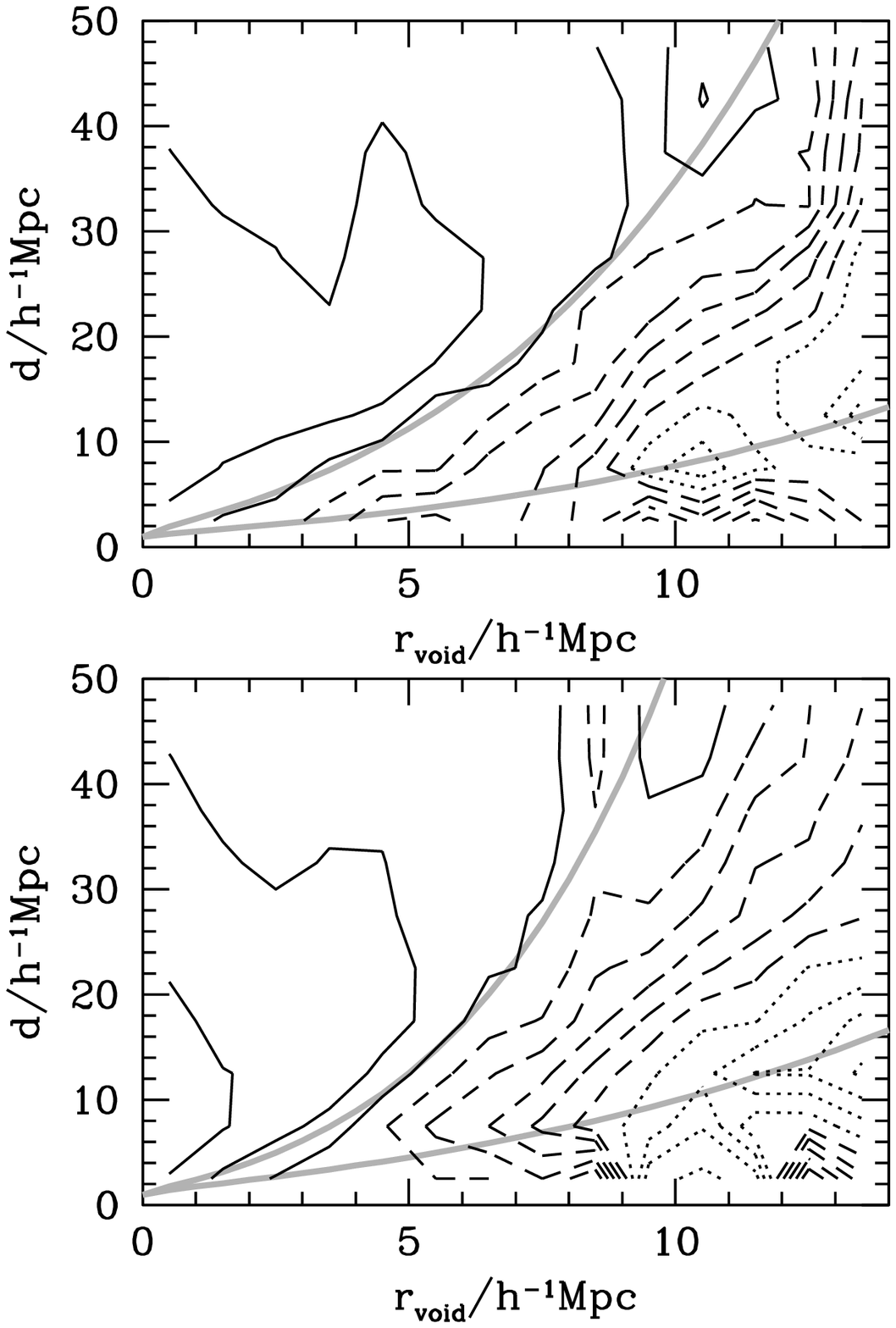,width=10.cm}}
\put(215,-20){\psfig{file=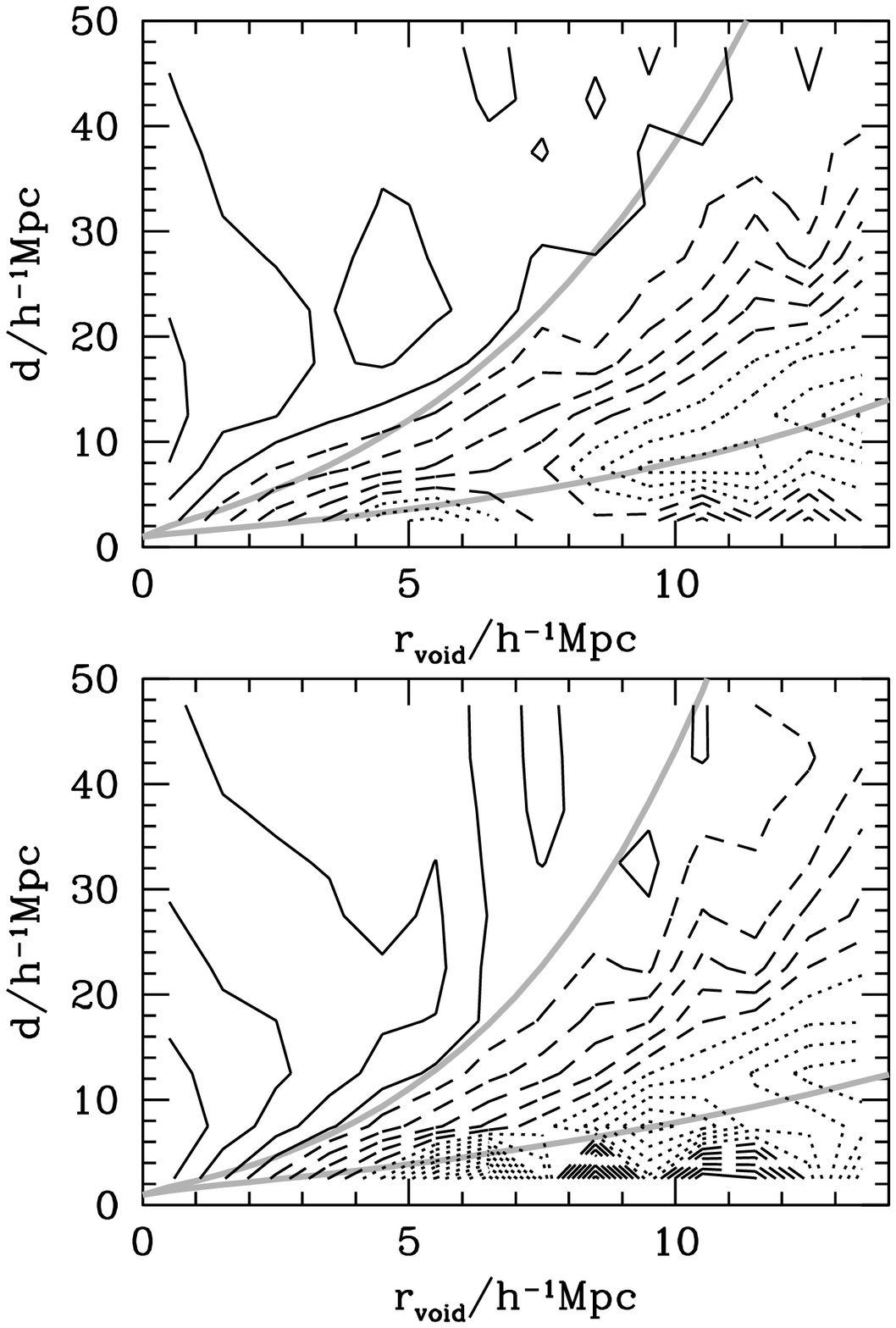,width=10.cm}}
\end{picture}
\caption{
2-dimensional outflow diagram of galaxies 
(top) and haloes ($M>10^{11.5}$h$^{-1}$M$_{\sun}$,
bottom) for galaxy- and halo-defined voids identified
from real- and redshift-space data (left and right respectively).
The x-axis corresponds to voidsize, the y-axis to distance from
the void centre.  Solid lines show inflow velocities 
$v_{\rm out}/$kms$^{-1}=0,20,40,...$
(positive velocities),
dashed lines show outflow motions, $v_{\rm out}/$kms$^{-1}=-20,-40,-60,-80$ and $-100$
(negative velocities), and 
dotted lines show stronger outflows, $v_{\rm out}/$kms$^{-1}=-120,-140,$ and so forth.
The thick gray solid lines show fits from eq. \ref{eq:vfit} 
for $d_{\rm vmin}$ and $d_{\rm zero}$ (lower and upper
lines respectively).  The best fit parameters are shown in table \ref{table:fits}.
}
\label{fig:outflow2d}
\end{figure*}

\begin{figure*}
\begin{picture}(430,250)
\put(0,-20){\psfig{file=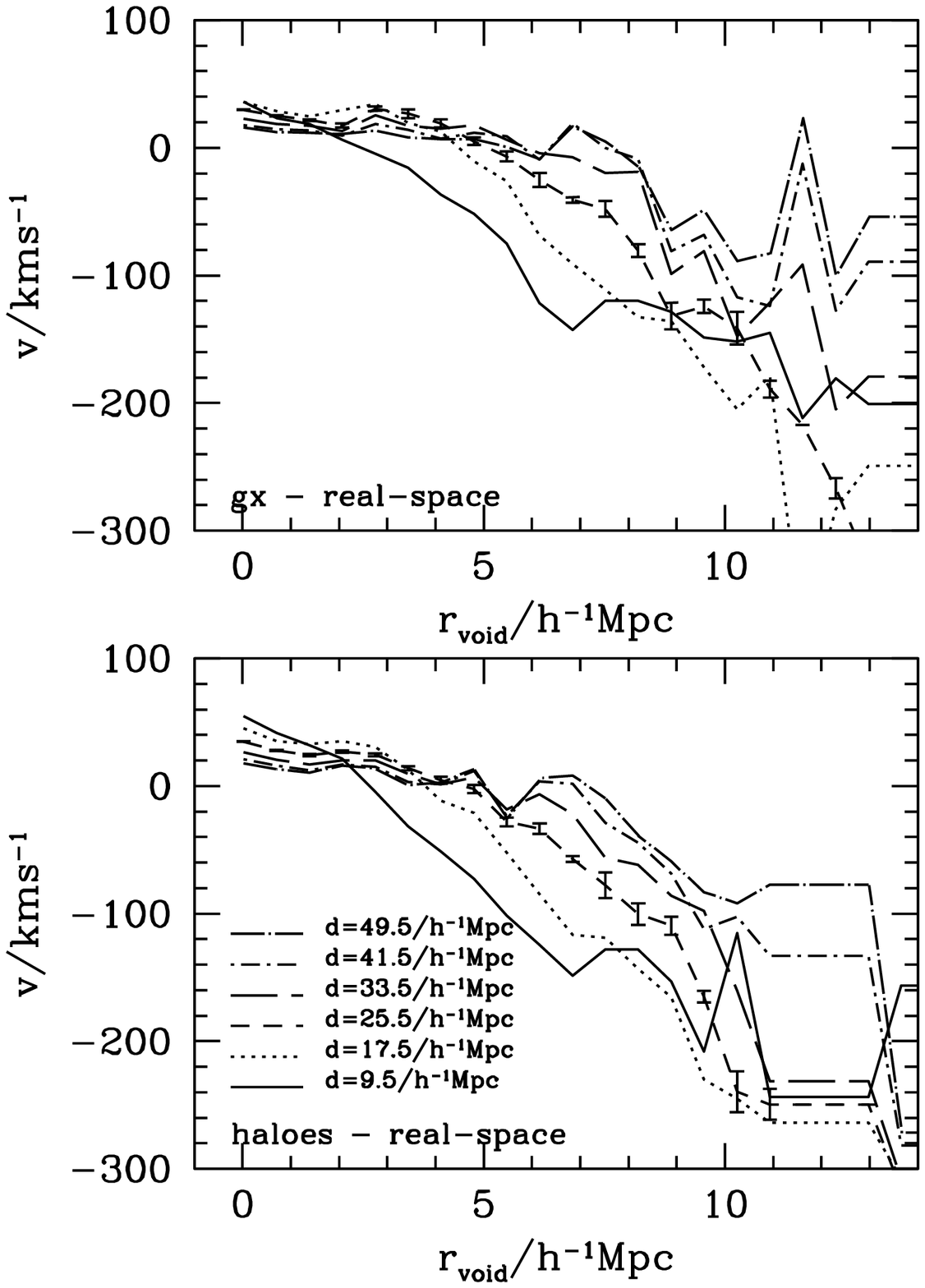,width=10.cm}}
\put(215,-20){\psfig{file=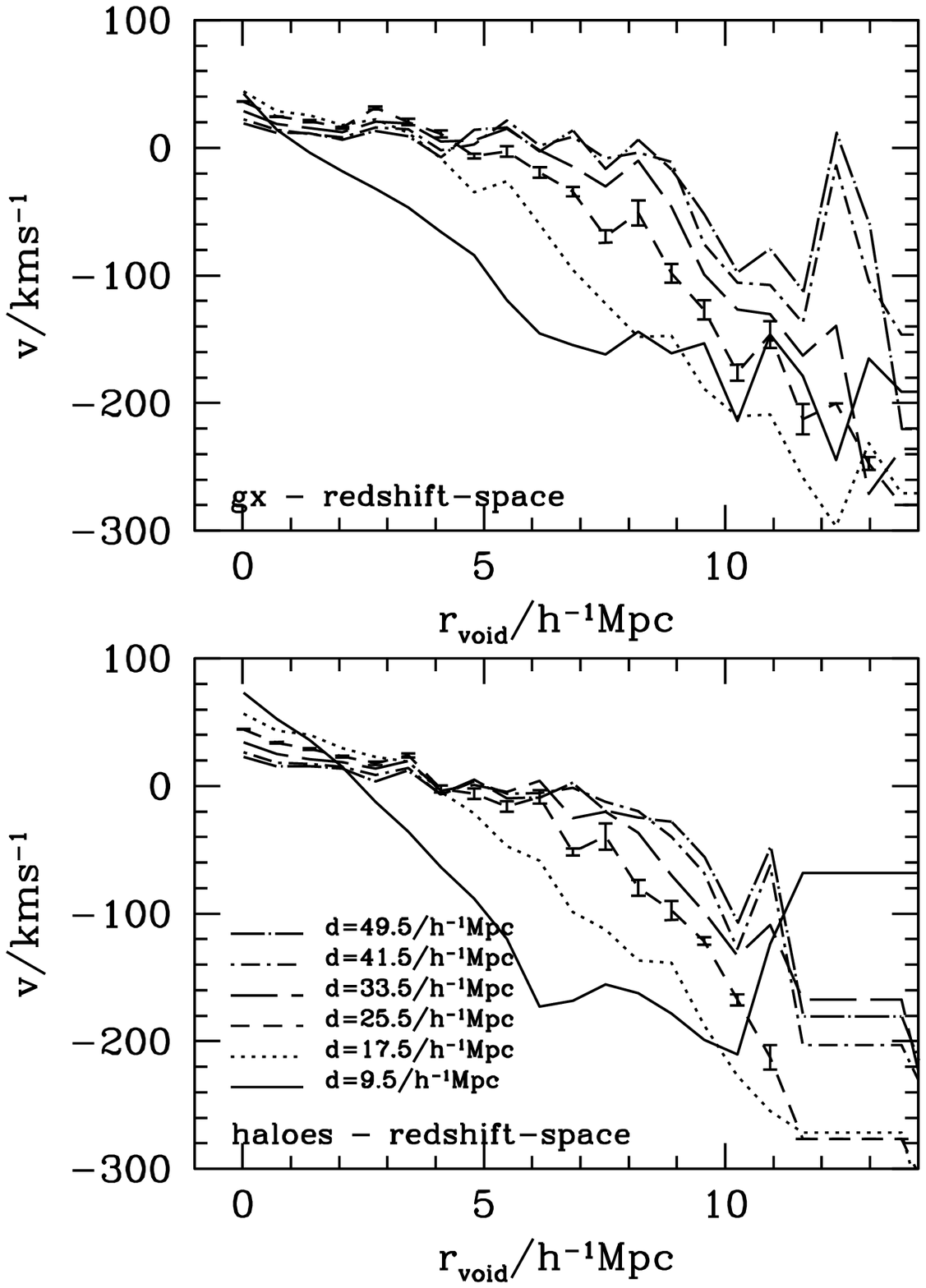,width=10.cm}}
\end{picture}
\caption{
Outflow velocities as a function of void radius, $r_{\rm void}$, 
of galaxies (top) and haloes ($M>10^{11.5}$h$^{-1}$M$_{\sun}$, bottom) around 
galaxy- and halo-defined voids respectively, 
identified from real- and redshift-space data (left and right panels 
respectively).  In order to improve clarity, jackknife
errorbars are only shown for $d=25.5$h$^{-1}$Mpc.
Ranges in $d$ are shown in the key.
}
\label{fig:outflowr}
\end{figure*}

\begin{figure*}
\begin{picture}(430,270)
\put(0,-10){\psfig{file=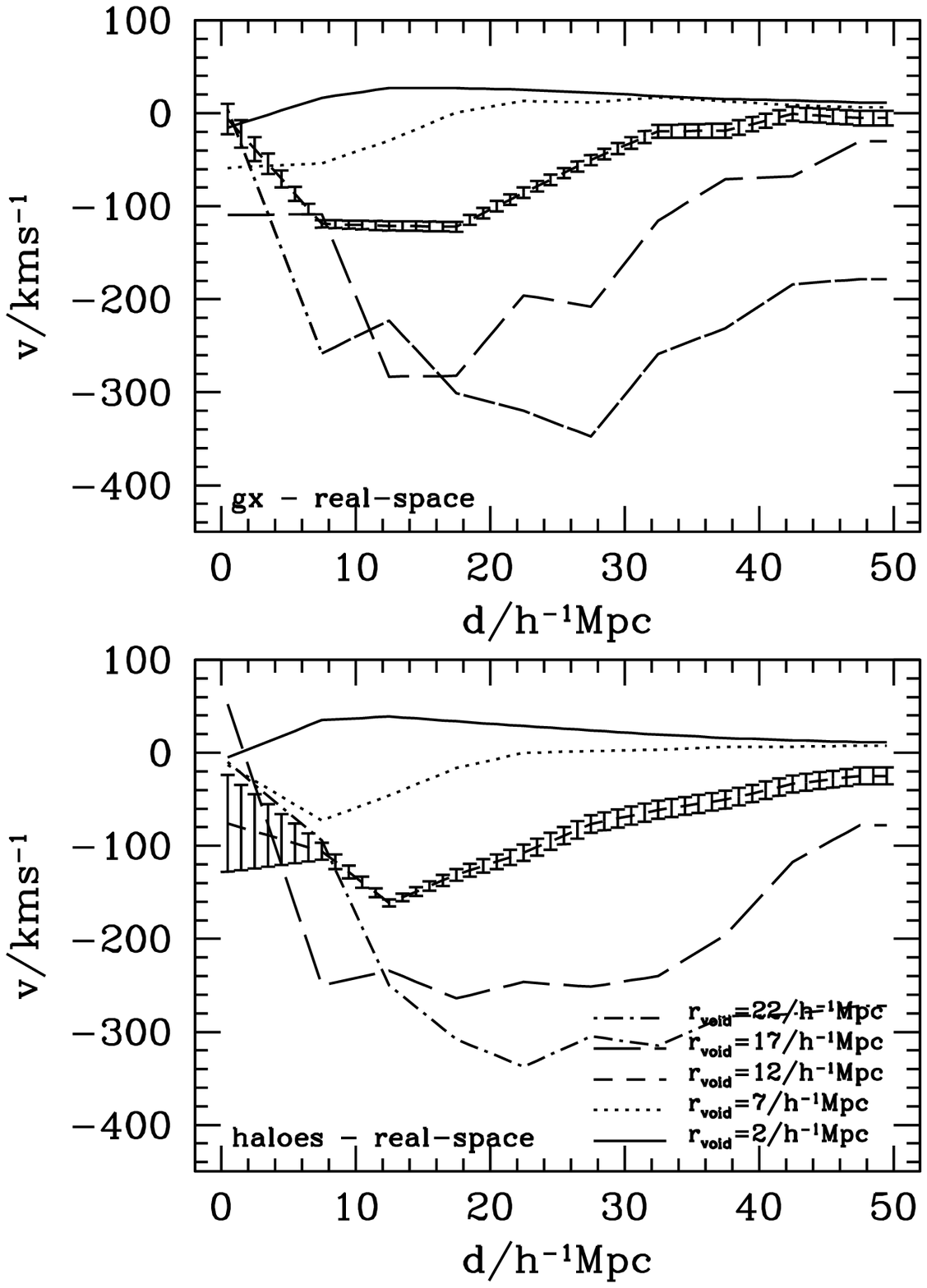,width=10.cm}}
\put(215,-10){\psfig{file=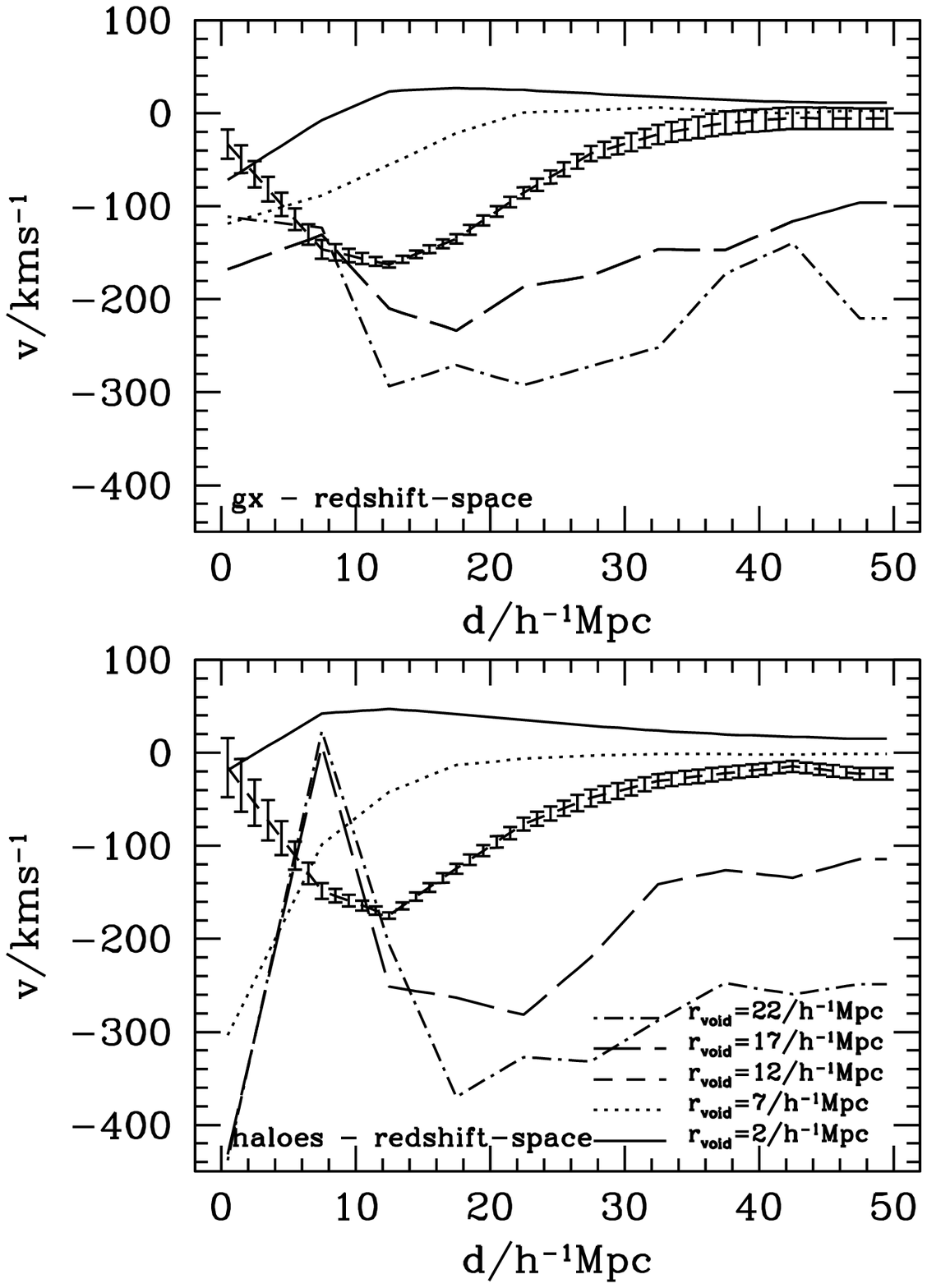,width=10.cm}}
\end{picture}
\caption{
Outflow velocities as a function of distance from
the void centre 
for galaxies (top) and haloes (bottom), around galaxy- and halo-defined
voids identified
from real- and redshift-space data (left and right panels 
respectively).  
In order to improve clarity, 
jackknife errorbars are shown only for $r_{\rm void}=12$h$^{-1}$Mpc.
Ranges in $r_{\rm void}$ are shown in the key.
}
\label{fig:outflowd}
\end{figure*}

\section{The peculiar velocity field around voids}
\label{sec:vpecs}

Just as it has recently become possible to study the peculiar velocity
field around groups (Ceccarelli et al. 2004b, C05 from now on), 
which is characterised
by infall galaxy motions toward the group centres, it is also possible
to study outflows of material escaping the void volumes.  A further
motivation to study the outflow motions is that as has been shown
by C05, the infall motions of mass and galaxies are almost identical, being
the galaxy infalls only slightly higher than the mass infalls.  It may
be possible then, to obtain a reliable measurement of the bulk outflows 
of mass emptying the void volumes by studying the velocities of galaxies
around voids.

We set out to compare the outflows of mass, dark-matter haloes, and galaxies from the
simulated void catalogues constructed in this work.  In order to do this we simply project
the peculiar velocity of galaxies and dark-matter haloes and particles onto
the radial direction measured towards the void centres.  
Adding the contribution from all the individual particles, haloes or galaxies together,
and averaging the result, we obtain a dependence of the outflow 
motions as a function of distance to the void, $d$, and void radius, $r_{\rm void}$.  
Defined in this way, a positive velocity indicates an infall to-wards the void centre,
and a negative velocity, an outflow.

Figure \ref{fig:outflow2d} shows the 2-dimensional diagrams of outflow
velocities as a function of void size (x-axis) and distance (y-axis).
The lines show constant outflow velocity levels; solid lines correspond
to positive velocities (infalls), and dashed and dotted lines show
negative velocities or outflows (see the figure caption for velocity level
values).  Lower panels in this figure show results for halo
velocities around H$11.5$-defined voids.  Upper panels show results for galaxy 
velocities around galaxy-defined voids.  As can be seen 
from the comparison of the iso-velocity contours in the upper
and lower panels, the velocities around halo-defined voids reach higher
values of both, infall and outflow motions, than around galaxy-defined voids by 
approximately $20$kms$^{-1}$ (see also figure \ref{fig:outflowd}).  However, if we
concentrate on the velocity minimum for large $r_{\rm void}$ voids,
we see that there is an agreement between results from haloes and galaxy-defined voids.
Also, small voids, $r_{\rm void}\leq7$h$^{-1}$Mpc, show infall motions at 
separations $d(r_{\rm void})\geq 20$h$^{-1}$Mpc.  
This indicates the existence of overdense regions
surrounding small voids, which would be responsible for producing the
observed infall motions.

We provide a fit for the distance at which
the outflow motions reach $v=0$kms$^{-1}$,
\begin{equation}
\log(d_{\rm zero}^{mass,r})= \left( \frac{r_{\rm void}}{A}\right)^B.
\label{eq:vfit}
\end{equation}
\noindent 
In the case of H$11.5$ velocities and H$11.5$-defined voids we find $A=3.0$h$^{-1}$Mpc and $B=0.6$, and for
galaxy-defined voids and galaxies, $A=3.5$h$^{-1}$Mpc and $B=0.7$.  
(for values of the parameters $A$ and $B$ for
the other simulated samples, and in redshift-space,
see table \ref{table:fits}). 
At larger distances from the void centre, velocities become positive, 
indicating an opposite infall motion.

The distance from the void centre where the minimum in outflow velocity 
occurs, $d_{\rm vmin}$, is very well approximated by
\begin{equation}
d_{\rm vmin}=r_{\rm void}.
\end{equation}
Motivated by the results for $d_{\rm zero}$, we can also fit eq. \ref{eq:vfit} 
to $d_{\rm vmin}$.  This equation could prove to be more accurate for 
larger voids.
In the case of H$11.5$ velocities in real-space, the best fit parameters
are $A=2.5$h$^{-1}$Mpc and $B=0.6$.  For galaxy-defined voids and galaxies, 
$A=3.6$h$^{-1}$Mpc and $B=0.7$.  The best fit parameters for all the samples studied here can 
be found in table \ref{table:fits}.  

The fits corresponding to $d_{\rm vmin}$ and $d_{\rm zero}$ 
as a function of the void radius
are shown in all panels of figure \ref{fig:outflow2d} in thick solid gray
lines (lower and upper lines respectively).

The fit parameters corresponding to
different halo masses correlate reasonably well with halo mass, and
can therefore be used to interpolate for samples of masses different
than the ones used in this paper.  Alternatively, this can also be
used when analysing a sample of galaxies for which the masses of
haloes in which they reside are known.  This is also applicable to
other fits provided in this work (Equations \ref{eq:vmin} 
and \ref{eq:rho}).

\begin{figure}
\begin{picture}(230,240)
\put(0,-10){\psfig{file=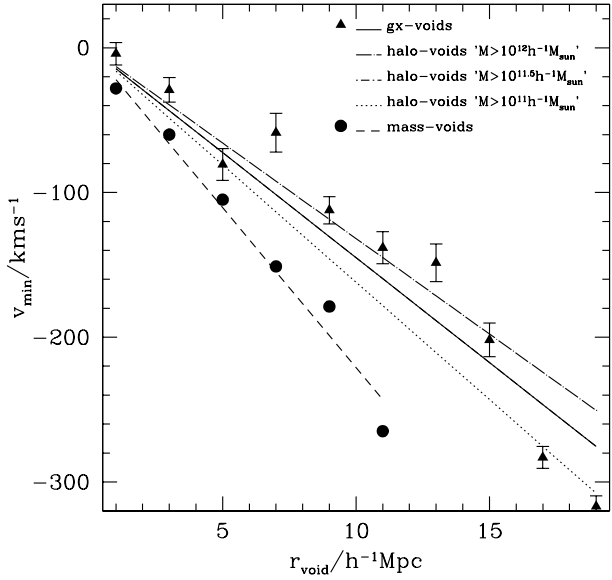,width=8.5cm}}
\end{picture}
\caption{
Maximum outflow velocity as a function of void radius
for galaxies (triangles) and dark-matter particles
(blue/black circles), around galaxy- and mass-defined 
voids identified from real-space data.
The lines show the linear fits to the points, $v_{\rm min}=v_0 r_{\rm void}$
(dashed and solid lines).  Dotted, dot-dashed, and long-dashed lines
correspond to linear fits to the minimum velocity found for
H$11$-, H$11.5$-, H$12$-defined voids.
Errorbars show the 1-$\sigma$ confidence errors calculated using the
jackknife method.
}
\label{fig:outflowmin}
\end{figure}

\begin{figure}
\begin{picture}(230,330)
\put(0,-10){\psfig{file=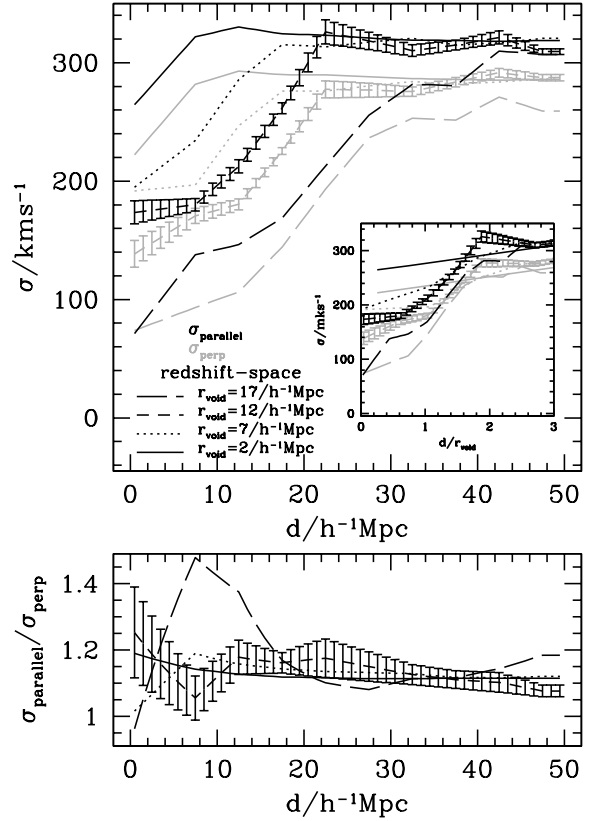,width=12.cm}}
\end{picture}
\caption{Upper panel:
Velocity dispersions in the directions parallel (black lines) and
perpendicular (gray lines) to the void walls for galaxy-defined voids,
as a function of distance to the void centre.
Different line types correspond to different void radii.  Error-bars
are only shown for the $r_{void}=12$h$^{-1}$Mpc for clarity, and are
calculated using the Jacknife method.
The inset shows the velocity dispersions as a function of normalised
distance to the void centre.
Lower panel: ratio between velocity dispersions parallel and perpendicular
to the void walls.  Line types are as in the upper panel.
}
\label{fig:wallv}
\end{figure}

\begin{table}
\caption{\small
{
Best fit parameters, $A$ and $B$, for the relation between void radius and
the distance at which the maximum outflow is measured ($d_{\rm vmin}(r_{\rm void})$),
and for the distance at which $v=0$ ($d_{\rm zero}(r_{\rm void})$).
We present the best fit parameters for voids selected from the distribution
of mass and galaxies, and in real- and redshift-space.
}}

\begin{tabular}{ccccc}
\hline
\hline
\noalign{\vglue 0.2em}
Statistic & Real/Redshift-space & Gx./mass & $A$hMpc$^{-1}$ & $B$\\
\noalign{\vglue 0.2em}
\hline
\noalign{\vglue 0.2em}
 $d_{\rm vmin}$ & Real     &  Mass  &  2.30 & 0.80\\
                &          &  H11   &  2.50 & 0.60\\
                &          &  H11.5 &  2.50 & 0.60\\
                &          &  H12   &  2.50 & 0.60\\
                &          &  Gx.   &  3.60 & 0.70\\
                & Redshift &  Mass  &  3.00 & 0.80\\
                &          &  H11   &  2.80 & 0.60\\
                &          &  H11.5 &  3.00 & 0.60\\
                &          &  H12   &  3.00 & 0.50\\
                &          &  Gx.   &  3.50 & 0.70\\
 $d_{\rm zero}$ & Real     &  Mass  &  0.10 & 0.38\\
                &          &  H11   &  1.00 & 0.65\\
                &          &  H11.5 &  1.20 & 0.65\\
                &          &  H12   &  1.70 & 0.65\\
                &          &  Gx.   &  1.00 & 0.55\\
                & Redshift &  Mass  &  0.10 & 0.37\\
                &          &  H11   &  1.00 & 0.65\\
                &          &  H11.5 &  1.30 & 0.65\\
                &          &  H12   &  1.70 & 0.65\\
                &          &  Gx.   &  0.95 & 0.55\\
\noalign{\vglue 0.2em}
\hline
\hline
\end{tabular}\label{table:fits}
\end{table}

We now focus on the dependence of the outflow velocities on 
void radius at fixed distances from the void centres.  We show
this dependence in figure \ref{fig:outflowr} where the left panels
show results in real-space and right panels in redshift-space.
Upper panels correspond to galaxies and galaxy-defined voids, and lower
panels to H$11.5$ haloes and H$11.5$-defined voids.  The different line types show
the outflow velocities at different distances from the void centres
(the key in the figure specifies their values).  The outflow
velocities at different distances show a complicated variation
as we go from regions near the void centres to large distances.  For instance,
it is easy to find mild infall motions $v<50$kms$^{-1}$ near
the centres of moderate sized voids with $r_{\rm void}<5$h$^{-1}$Mpc. 
Also, the smaller the distance to the void centre, the smaller the void
at which an outflow motion (negative velocities) can be found.
The trend of velocities around voids with increasing void radius
is negative for $r_{\rm void}\geq 5$h$^{-1}$Mpc.

A much simpler dependence characterizes the outflow velocity as a function
of distance for different void sizes, as can be seen in figure
\ref{fig:outflowd}.  In this figure, the left- and right-hand panels
show results in real- and redshift-space respectively; upper panels
show results from the galaxy distribution, lower panels from the
H$11.5$ distribution.  The different line types in these plots correspond
to different void radii.  In general, it can be seen that 
larger voids reach a minimum
in velocity (or a maximum outflow) at larger separations.  The outflow is
also stronger for larger voids.  
At larger distances from void centres, velocities tend to $ v=0$ as
expected given the lack of correlation of structure at these scales.
Still, it should be 
noted that the outflow signal continues to be present at distances larger 
than the void size.  

Figure \ref{fig:outflowd} also shows that smaller voids 
($r_{\rm void}<10$h$^{-1}$Mpc) show a rapid transition in the very nearby 
regions, going from outflow to infall motions, revealing that these may be 
immersed in the middle of high density regions responsible for the infalls.
Note however, that the amplitude of such infall motions is small 
compared to the infall motions around groups of galaxies (C05).  Another 
contribution to this infall could originate in outflows from neighbouring voids.

We can also find out about the differences arising
from identifying voids in the distribution of galaxies and dark-matter haloes
in figure \ref{fig:outflowd}.  In 
this visualization, the galaxy-defined voids also tend to show slightly stronger outflows
than the H$11.5$-defined voids, and small galaxy-defined voids show a smaller degree of infall
than their H$11.5$ counterparts.  
The effects of redshift-space selection
simply tend to decrease the range of velocities measured around voids, that is,
infall and outflow extremes are lower in absolute value, both for
galaxy and H$11.5$-defined voids.  
These effects must be taken into account when comparying observational data
on the dynamics of galaxies around voids with the corresponding theoretical predictions.

\begin{figure}
\begin{picture}(200,400)
\put(0,-20){\psfig{file=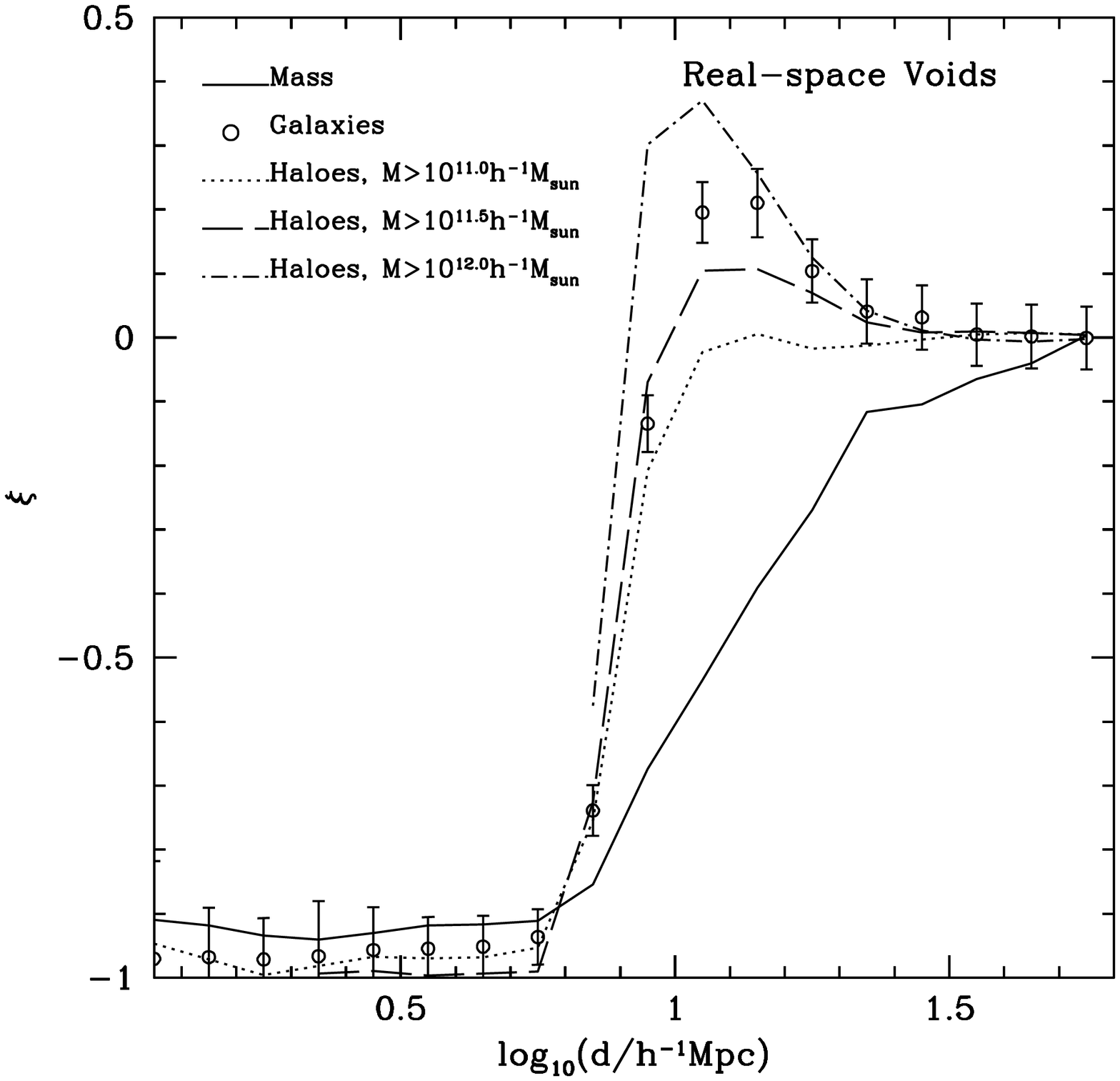,width=8.cm}}
\put(0,190){\psfig{file=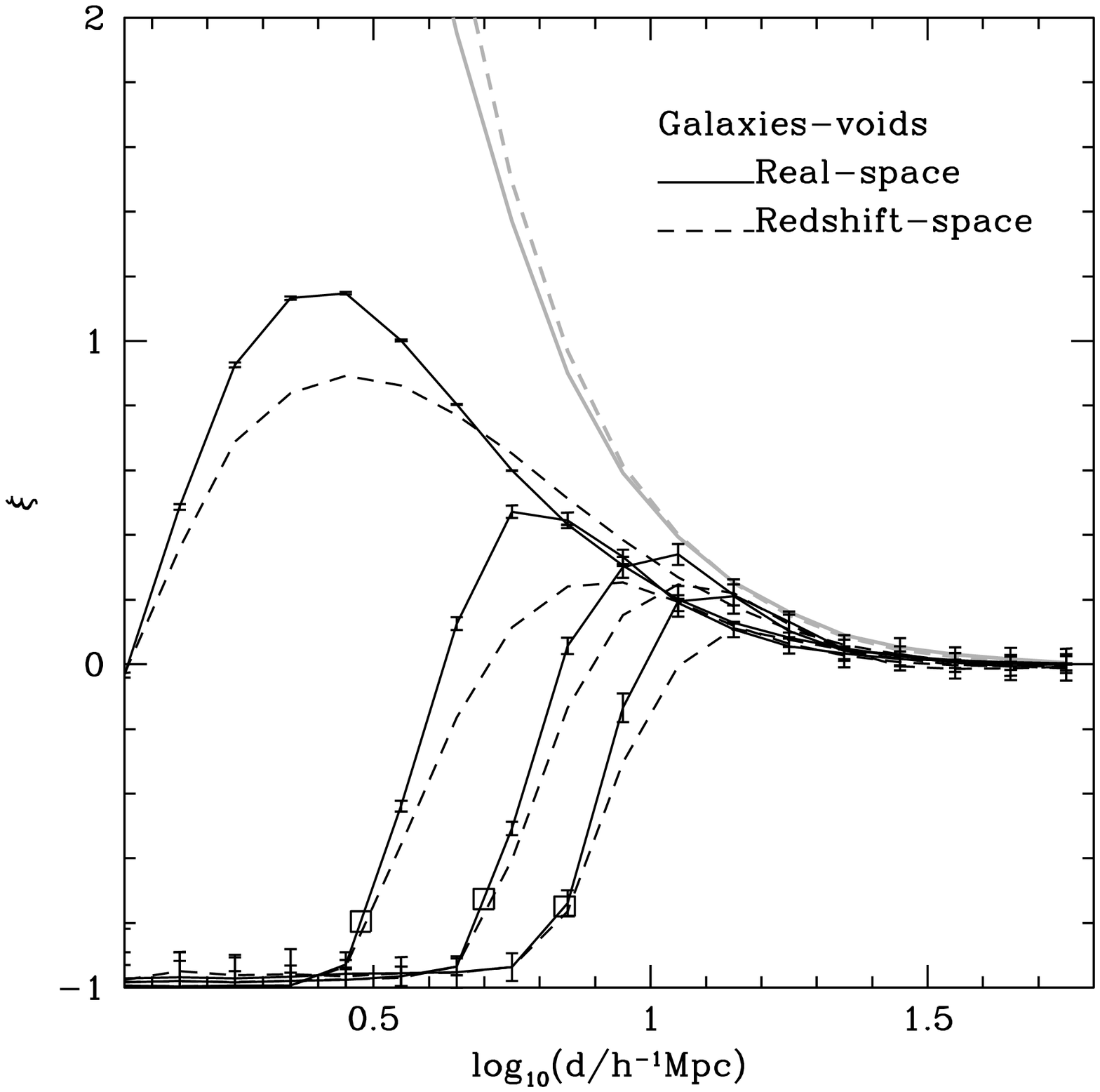,width=8.cm}}
\end{picture}
\caption{
Upper panel:
Real- and redshift-space cross correlation functions
(solid and dashed lines respectively) between
centres of galaxy-defined voids identified in real- and redshift-space
respectively, and galaxies in the simulation.  From
left to right, the black solid and dashed lines
represent the results for $0<r_{\rm void}/$h$^{-1}$Mpc$<2$,
$2<r_{\rm void}/$h$^{-1}$Mpc$<4$, $4<r_{\rm void}/$h$^{-1}$Mpc$<6$, and
$6<r_{\rm void}/$h$^{-1}$Mpc$<8$.  The squares
show the correlation function level at $d=r_{\rm void}$.
Grey curves show the galaxy auto-correlation
functions in real- (solid) and redshift-space (dashed lines).
Lower panel: comparison between cross-correlations
around mass-, H$11$-, H$11.5$-, H$12$-, and galaxy-defined voids
(solid, dotted, long-dashed, dot-dashed lines and circles with errorbars,
respectively) for $8<r_{\rm void}/$h$^{-1}$Mpc$<12$,
in real-space.
}
\label{fig:xis}
\end{figure}

\begin{table}
\caption{\small
{
Best fit parameter values for the 
maximum outflow around voids, as a function of void radius.
We present the best fit parameters for voids selected from the distribution
of mass, haloes and galaxies, in real-space.
}}

\begin{tabular}{cc}
\hline
\hline
\noalign{\vglue 0.2em}
Gx./halo,mass &  $v_0$km$^{-1}$sh$^{-1}$Mpc\\
\noalign{\vglue 0.2em}
\hline
\noalign{\vglue 0.2em}
Mass &  -22.1 \\
H11  &  -16.2 \\
H11.5&  -14.5 \\
H12  &  -13.2 \\
Gx.  &  -14.5 \\
\noalign{\vglue 0.2em}
\hline
\hline
\end{tabular}\label{table:vmin}
\end{table}
Figure \ref{fig:outflowmin} shows the dependence of the maximum
outflow velocity with void radius.  As can be seen, the
dependence is well approximated by a linear relation 
with the void radius.  The best fit
parameters of the relation
\begin{equation}
v_{\rm min}=v_0 r_{\rm void}
\label{eq:vmin}
\end{equation}
can be found in table \ref{table:vmin} for the different choices
shown in the figure.  It can be seen that
mass-defined voids show a much lower and steeper maximum velocity relation
when compared to galaxies,
which show a difference of $\simeq 150$kms$^{-1}$ at
$r_{\rm void}\simeq 10/$h$^{-1}$Mpc with respect to the results from 
mass-defined voids.  In this case, it is not
possible to decide which sample of halo-defined voids provides
the closest outflow velocities to the galaxies,
since the data from each halo sample
are consistent with the galaxy-void
data (which shows considerable scatter around the best fit line).  
Still, the results from H$11.5$-defined voids are characterised
by the same best fit $v_0=-14.5$kms$^{-1}$hMpc$^{-1}$ than the galaxy-void data.

We have also investigated 
the effect of using redshift-space positioning
and found that the linear relation between $v_{\rm min}$ and void radius
steepens considerably $\Delta v_0<-5$kms$^{-1}$hMpc$^{-1}$,
independently of the sample of voids used.

Finally, we also studied the possible differences in
the velocity field in the directions parallel and perpendicular
to the void walls.  In order to do this we calculated the 1-D velocity
dispersion of galaxies in both directions ($\sigma_{parallel}$ and
$\sigma_{perp}$, respectively), removing the local outflow
velocity when calculating the radial velocity
dispersion.  Our findings are summarised in the upper panel of 
figure \ref{fig:wallv},
where the black lines show $\sigma_{parallel}$ and the gray
lines show $\sigma_{perp}$.  Different line types correspond to
different void radii as indicated in the key, and we only show
results for voids identified using galaxies in redshift-space since
there is little variation when using haloes or real-space positions.
As can be seen, the velocities aligned with the radial direction from
the void centre are about $10-20\%$ lower than the velocities on the direction
of the void walls, regardless of void size and distance to void centre,
indicating that galaxies have a slight but systematic 
($2-$ to $3-\sigma$ level) tendency to follow trajectories 
along the void surfaces.  
The inset in the upper panel shows the dependence of the velocity
dispersions in the directions parallel and perpendicular to the line
of sight as a function of the distance to the void centre normalised 
by the void radius.  As can be seen, at $d/r_{\rm void}\simeq2$,
the curves corresponding to different void radii converge.  On the other hand,
smaller voids show much higher velocity dispersions at $d/r_{\rm void}<2$ when
compared to larger voids.  This is in agreement with the indication that
small voids are embedded in higher density regions, which could be respondible
for the high velocity dispersions in the small voids inner regions.
The lower panel of this figure shows the ratio
between the velocity dispersions in the directions parallel and perpendicular
to the void walls, and as can be seen there is little dependence on the distance
to the void centre and on the void radius, being consistent with a roughly
constant $\sigma_{parallel}/\sigma_{perp}=1.1$.

\vskip .5cm

The observational study of the peculiar velocity field around voids 
is carried out in a forthcoming paper, Ceccarelli et al. (2004),
where the different observational biases present in peculiar
velocity data are analysed, and the results are compared
to the theoretical expectations presented here.

\section{Correlations around voids}
\label{sec:correlations}

We now focus our attention on the spatial statistics of
galaxies, haloes and mass around voids.  We use the 2-point
cross correlation function between void centres, and
galaxies, haloes or dark-matter particles to characterise the
spatial clustering properties of the vicinities of voids.
In the following subsections we study the 1-dimensional
cross-correlation function in real- and redshift-space,
and the cross-correlation function as a function of the
coordinates perpendicular and parallel to the line of sight.

\subsection{1-Dimensional correlation functions}

In this subsection we study the spatial correlations
between the void centres and the objects surrounding them.  We will
compare once more the resulting statistics from considering
voids in the distribution of galaxies, haloes and mass, using
real- and redshift-space positions.

We start by analysing our expectations for the outcome of a void-object
cross-correlation function, which we compute using the following
estimator, 
\begin{equation}
\xi(r)=\frac{d_cd_o}{d_cr_o}-1,
\end{equation}
where $d_cd_o$ corresponds to the number of center-object pairs,
and $d_cr_o$ is the number of center-random pairs which would 
be measured if the positions of objects were uniformly distributed
in space.  The latter is calculated using the mean density of
objects in the numerical simulation box.
What we expect to find at small separations is an anti-correlation,
lower in amplitude than the maximum overdensity set for identifying voids,
that is, $\xi<-0.9$.  On the other hand, at sufficiently large 
separations, $d\simeq r_{\rm void}$,  we would expect to
find departures from this value and to gradually tend towards
a small but positive value for the correlation function, indicating a population
of galaxies at the void walls.  We also expect $\xi\simeq0$
at sufficiently large separations from the void centres.

Figure \ref{fig:xis} shows the cross-correlation functions between galaxy-defined voids 
and galaxies (upper panel), and galaxy-, halo- and mass-defined voids with their
respective counterparts, in real space (lower panel).
Solid and dashed lines in the upper panel
represent the results in real- and redshift-space respectively.  
Concentrating on the results for galaxy-defined voids, it can be seen that we 
obtain the expected value $\xi<-0.9$ at low
separations for void radii $r_{\rm void}>4$h$^{-1}$Mpc in real- and redshift-space
(Smaller voids may show this effect at lower separations than those analysed
in this work), and also that at the void radius, $\xi(r_{\rm void})\simeq-0.9$ (open squares).
This relation holds regardless of the tracer used to identify voids.  The 
general effect produced by changing from real- to redshift-space data,
where the voids are identified in redshift-space and the 
tracers are also in redshift-space, is 
that of lowering very slightly the maximum clustering amplitude.

Another important feature of the correlation functions shown in the 
upper panel of figure \ref{fig:xis} is that at separations, 
$d\simeq r_{\rm void}$, $\xi$ increases rapidly in value 
reaching $xi>0$, but always lower than 
the galaxy auto-correlation function.  It is also
noticeable the fact that the correlation function around
voids can be positive outside the void radius, particularly for
the smaller voids, $r_{\rm void}<4$h$^{-1}$Mpc, which can be interpreted
as small voids being embedded in extended, reasonably high  
density regions,
in accordance with the results from the velocity field around
small voids, which shows systematic infalls instead of outflows. 
Note that this peak in $\xi$ is still
quite low compared to the galaxy correlation function. 

From the
comparison between results for different void populations in the lower
panel of figure \ref{fig:xis}, we see that mass-defined voids
show a higher cross-correlation function than haloes and galaxies
at $d<r_{\rm void}$.  
Haloes, on the other hand, show the same correlation levels
than galaxies at small separations from the void centres.
At larger separations, $d>r_{\rm void}$, individual dark-matter particles are 
less clustered than galaxies.  Haloes show a range of
clustering amplitudes which correlates with the halo mass.
In particular, H$11.5$ show consistent correlation amplitudes
than galaxies, and H$12$ show higher $\xi$s.

It is interesting to note though, that independently of the object
used to identify voids,  the cross-correlation function
shows a similar qualitative shape,
starting off at a constant negative value at low separations, 
independent of void radius, and increasing
at separations of the order of the void radii, approaching $\xi=0$.
Note that this change from negative to $\xi\simeq 0$ is
sharper for haloes and galaxies than for mass-defined voids.
For instance, for $r_{\rm void}>8$h$^{-1}$Mpc, the galaxy-void - galaxy 
cross-correlation function values are as high as the galaxy $\xi$.  

\begin{figure}
\begin{picture}(210,570)
\put(0,-10){\psfig{file=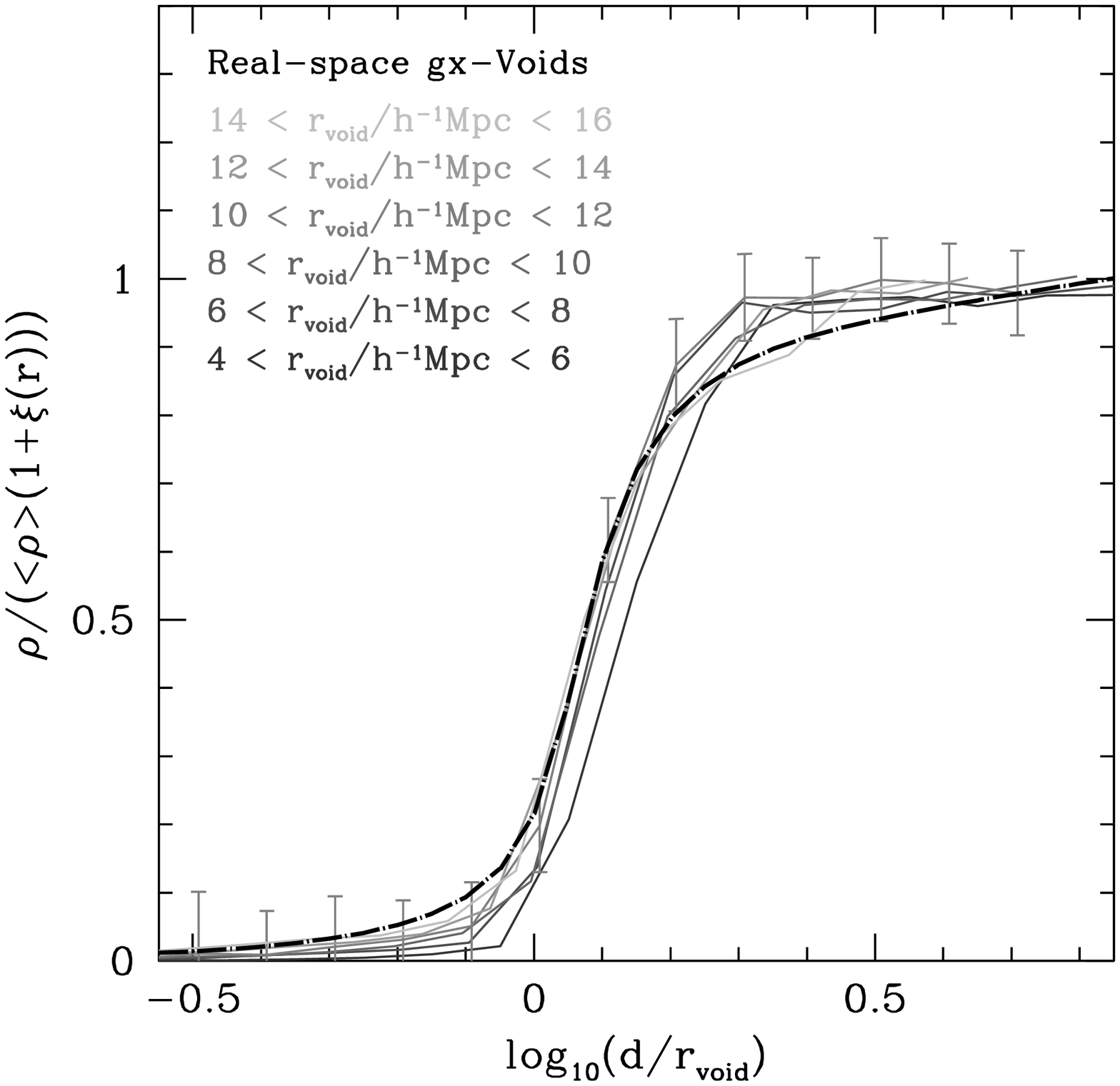,width=7cm}}
\put(0,180){\psfig{file=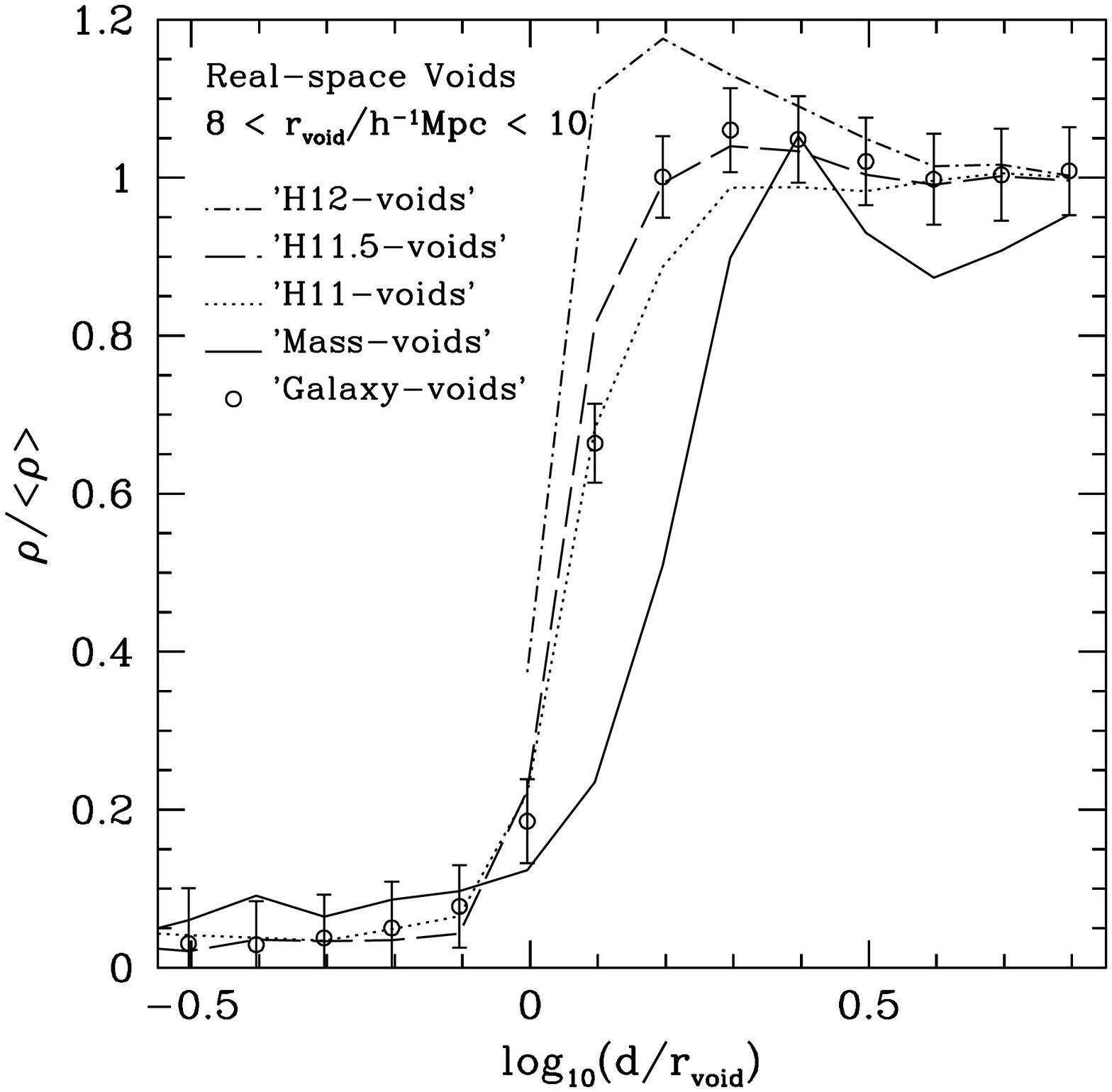,width=7.cm}}
\put(0,360){\psfig{file=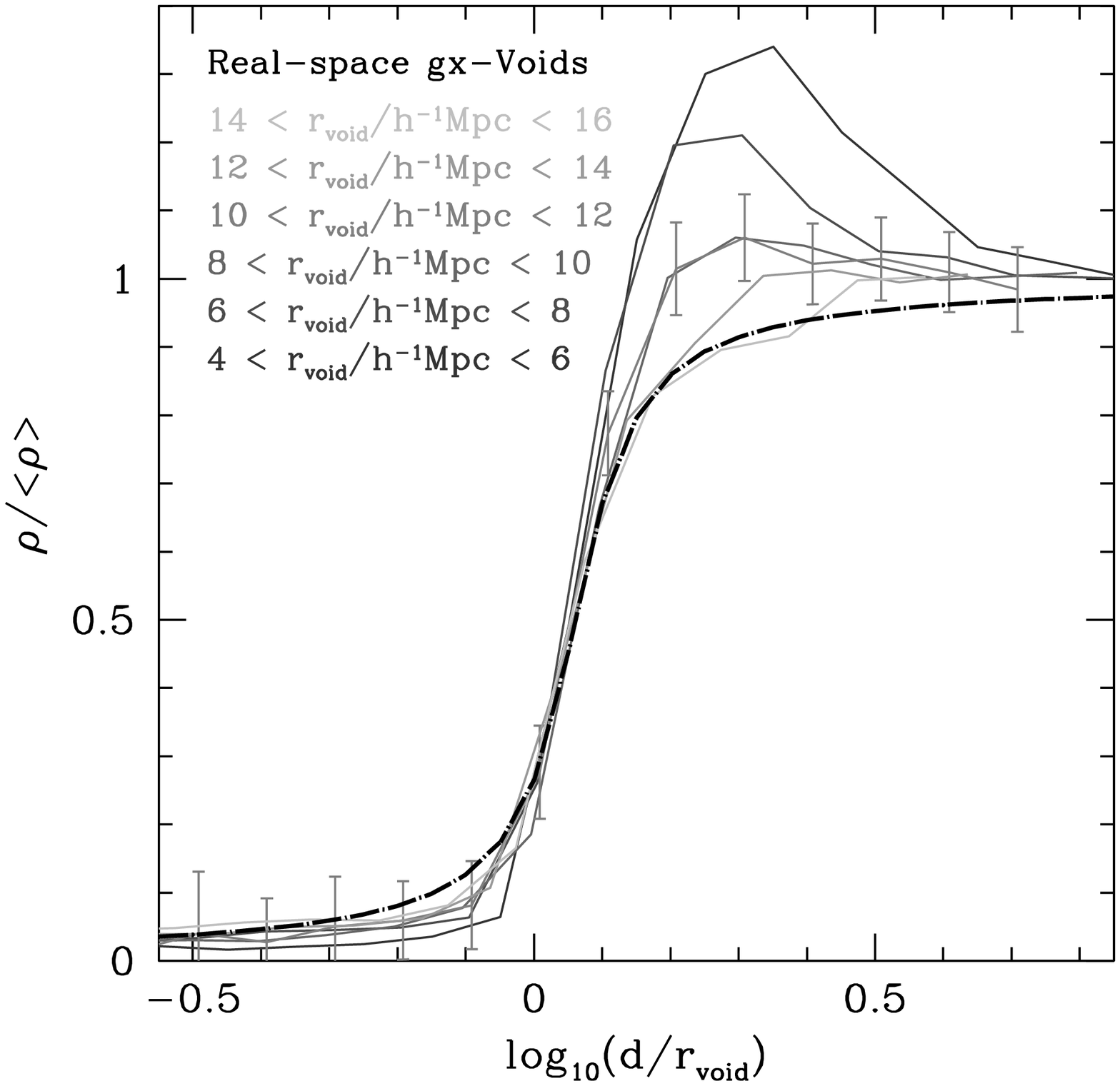,width=7cm}}
\end{picture}
\caption{
Upper panel: Real-space density profiles
around centres of galaxy-defined voids 
(circles and triangles respectively) in the simulation
as a function of reduced distance, $d/r_{\rm void}$.  
Gray-scale lines represent the results for
$8<r_{\rm void}/$h$^{-1}$Mpc$<10$, $10<r_{\rm void}/$h$^{-1}$Mpc$<12$, $12<r_{\rm void}/$h$^{-1}$Mpc$<14$,
and $14<r_{\rm void}/$h$^{-1}$Mpc$<16$ (lighter to darker gray-scale).  
The dot-dashed line shows the the empirical fit (see text) for the
galaxy-void density profile.  Lower panel: comparison between
the measured density profiles of 
mass-, H$11$-, H$11.5$-, H$12$-, and galaxy-defined voids
(solid, dotted, long-dashed, dot-dashed lines and circles with errorbars,
respectively) for $8<r_{\rm void}/$h$^{-1}$Mpc$<12$,
in real-space.
}
\label{fig:rho}
\end{figure}

Given the clear correlation between the void-size and the distance at 
which the correlation amplitude departs from its constant negative 
value, we have also calculated the density profiles around 
voids as a function of the reduced distance to the void
centres, $d/r_{\rm void}$ (figure \ref{fig:rho}).  As can be seen, 
density profiles for voids with $r_{\rm void}>8$h$^{-1}$Mpc are
marginally consistent with one another ($2-\sigma$). 

The comparison between the
density profiles of different void populations, shown in the middle panel
of figure \ref{fig:rho}, is similar to what was found for cross-correlation
functions.  Namely, we find that the density profiles of mass-defined voids 
are lower in amplitude than galaxy-defined voids.  The latter are found to
be of similar amplitudes than the density profiles of halo-defined voids,
in agreement with the numerical simulation results by Gotl\"ober et al. 2003.
On large
separations $d/r_{\rm void}>1$,
small voids with $r_{\rm void}<8$h$^{-1}$Mpc 
show density profiles much higher than
larger voids with $r_{\rm void}>8$h$^{-1}$Mpc.

The differences in $\rho/<\rho>$ for the different types of void
can be alleviated significantly by dividing the density profile by the correlation
function of the objects used to identify the voids (see the lower panel
of figure \ref{fig:rho}).  In this case, the density profiles
divided by the correlation function of galaxies are in better agreement
with one another for a wider range of void sizes, $r_{\rm void}>2$h$^{-1}$Mpc.
In addition, this ratii can be fitted by the 
empirical function,
\begin{equation}
\frac{\rho(r)}{<\rho>(1+\xi)}=1-A_1\exp \left[- \left( A_2 \frac{d}{r_{\rm void}}\right)^c\right].
\label{eq:rho}
\end{equation}
\noindent The parameter $A_1$ is related to the density contrast
threshold used for defining the voids ($\delta=-0.9$ in our case), 
and to whether galaxies, haloes or mass
are used as the tracers of the density field.  $A_2$ depends strongly on
the tracer and on the maximum density contrast used to identify
the voids.  The parameter $c$ indicates 
how rapid is the change from constant negative density contrast to the 
mean density of the Universe and also depends strongly on the tracer.
Table \ref{table:rho} summarises the parameter values corresponding to the
different choices we show in the figure.
The use of different void-finding algorithms will also change
the values of the quoted best-fit parameters.  For instance, the use
of the distance to the closest object (as in Gotl\"ober et al. 2003),
will lower the value of $A_2$ to compensate for the typically smaller
radii found by this void-finding method.

The thick dot-dashed line in the inset shows again the fit in
the main panel, this time divided by the galaxy correlation function; 
the $\chi^2$ per degree of freedom is
always lower than $1$ for $r_{\rm void}>4$h$^{-1}$Mpc in this case.  As a final check,
we have corroborated that this level of coincidence between
density profiles divided by correlation functions also holds for H$11$-,
H$11.5$-, H$12$-, and mass voids.

In sum, we find 
that voids with $r_{\rm void}>8$h$^{-1}$Mpc,
show a nearly 
universal radial profile for
the ratio involving the density profile of objects and the 2-point auto-correlation
function, $\rho/(<\rho>(1+\xi))$.  This is true for voids
identified from both, galaxies and dark-matter haloes.
The best fit parameters for the density profiles are
presented in table \ref{table:rho}.  
Since voids of any size (within the limits imposed by the
numerical simulation available for this work, $2<r_{\rm void}/$h$^{-1}$Mpc$<30$)
show similar radial $\rho/(<\rho>(1+\xi))$ profiles.  The parameters in the
table also describe this profile provided the fit is
divided by the correlation function of the tracers used when constructing
the sample of voids.

\begin{table}
\caption{\small
{
Best fit parameters, $A_1$, $A_2$ and $c$, for the density
profile of dark-matter particles, dark-matter haloes and galaxies,
around mass- halo- and galaxy-defined voids respectively, 
$\rho(r)/<\rho>=1-A_1\exp \left[ -\left( A_2 \frac{d}{r_{\rm void}}\right)^c\right]$.
}}

\begin{tabular}{ccccc}
\hline
\hline
\noalign{\vglue 0.2em}
Real/Redshift-space&Gx./mass & $A_1$ & $A_2$ & c\\
\noalign{\vglue 0.2em}
\hline
\noalign{\vglue 0.2em}
Real     & Mass &  0.90  & 0.60  & 5.0 \\
         & H11  &  0.96  & 0.80  & 5.5 \\
         & H11.5&  0.97  & 0.85  & 7.0 \\
         & H12  &  0.98  & 0.89  & 8.0 \\
         & Gx.  &  0.96  & 0.80  & 6.0 \\
Redshift & Mass &  0.90  & 0.65  & 5.0 \\
         & H11  &  0.95  & 0.85  & 5.5 \\
         & H11.5&  0.96  & 0.90  & 6.0 \\
         & H12  &  0.98  & 0.93  & 7.0 \\
         & Gx.  &  0.95  & 0.80  & 5.0 \\
      
\noalign{\vglue 0.2em}
\hline
\hline
\end{tabular}\label{table:rho}
\end{table}

\subsection{Redshift-space distortions}

We study the galaxy-void - galaxy and halo-void - halo cross-correlation functions, 
$\xi(\sigma,\pi)$,  as a function of the coordinates parallel ($\sigma$) 
and perpendicular ($\pi$) to the line of sight, which we take to be the
$z-$axis of the numerical simulation.  In order to simulate
the redshift-space distortions we again displace the positions of
dark-matter haloes and galaxies by the projection onto the $z$-axis
of their peculiar velocities and measure $\xi(\sigma,\pi)$.
In order to visualize this 2-dimensional distorted function, we study the
the contours of constant correlation amplitude in the
$\sigma-\pi$ plane.  We remind the reader that the iso-correlation contours for
a $\xi(\sigma,\pi)$ measured using real-space
positions would correspond to circles,
since there are no preferred directions in the simulation and the
correlation function must therefore be isotropic.  However, as we use redshift-space
positions, we can learn about the dynamical properties of galaxies,
dark-matter haloes and individual particles around voids from the departure of
contour shapes from perfect circles.  Such departures are
commonly known as redshift-space distortions.

Redshift-space distortions have been extensively studied for all kinds
of objects, including quasars (Hoyle et al. 2002), galaxies
(Loveday et al. 1995, Rattclife et al. 1998), groups (Padilla et al. 2001), 
and clusters of galaxies (Bahcall \& Cen 1992, Padilla \& Lambas 2003a and 2003b).
Also, assuming that redshift-space distortions around voids are not important,
Ryden (1995) proposed using the cosmological distortions around voids for
measuring $q_0$.

At small separations, the galaxy correlation function, $\xi(\sigma,\pi)$, shows elongations in 
the direction of the line of sight due to their random motions in 
the interiors of clusters of galaxies (Loveday et al. 1995).  At larger separations,
it is possible to see infall motions toward large mass concentrations
(Hawkins et al. 2003).  The effect of infall motions was first measured
using galaxy groups in the UZC by Padilla et al. (2001).  The
advantage of using groups relies in the fact that they 
only show infall signatures in their distortion patterns, which
induces a flattening of the correlation function contours in the direction of
the line of sight.  

\begin{figure*}
\begin{picture}(390,600)
\put(0,-50){\psfig{file=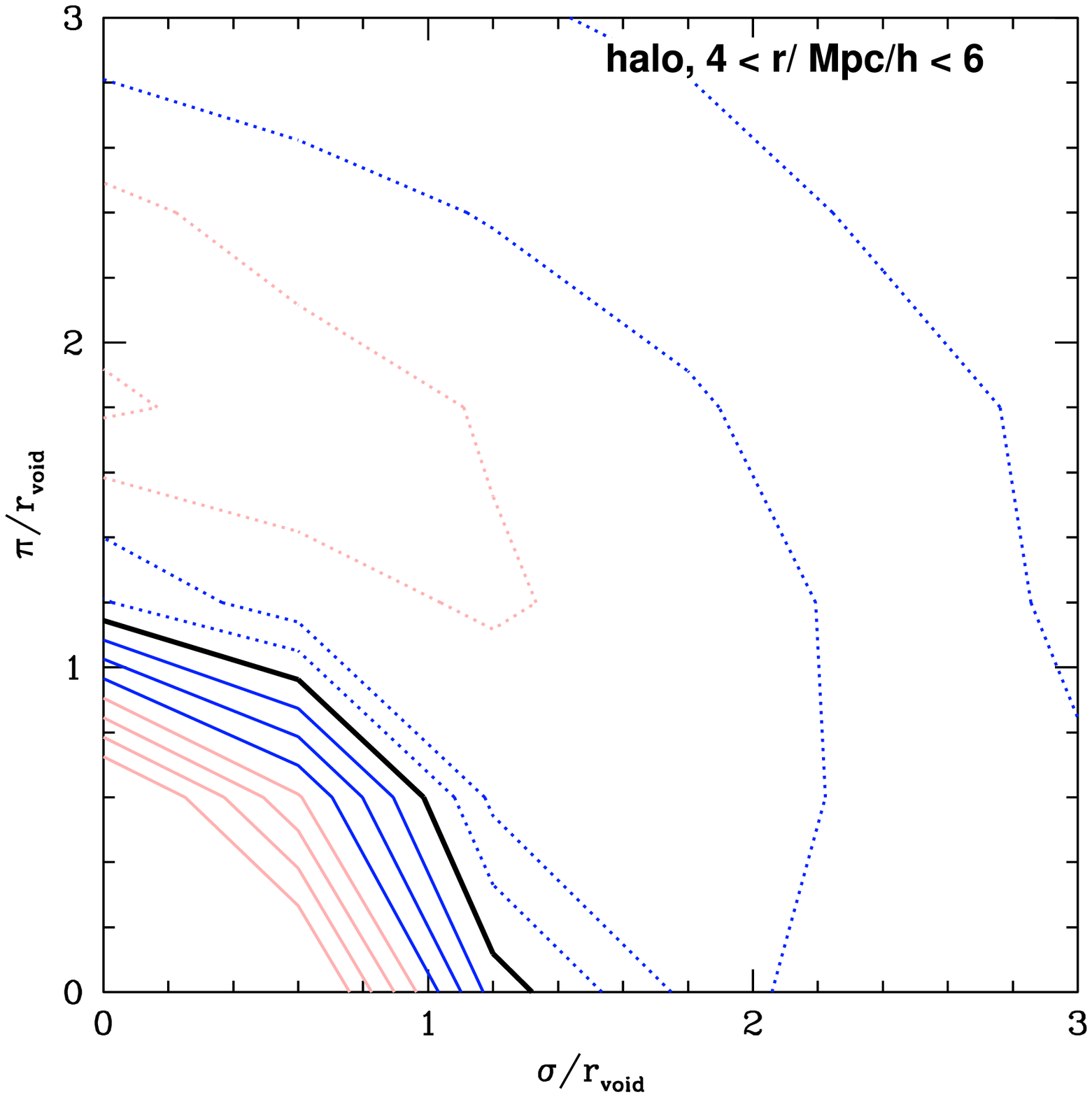,width=7.cm}}
\put(195,-50){\psfig{file=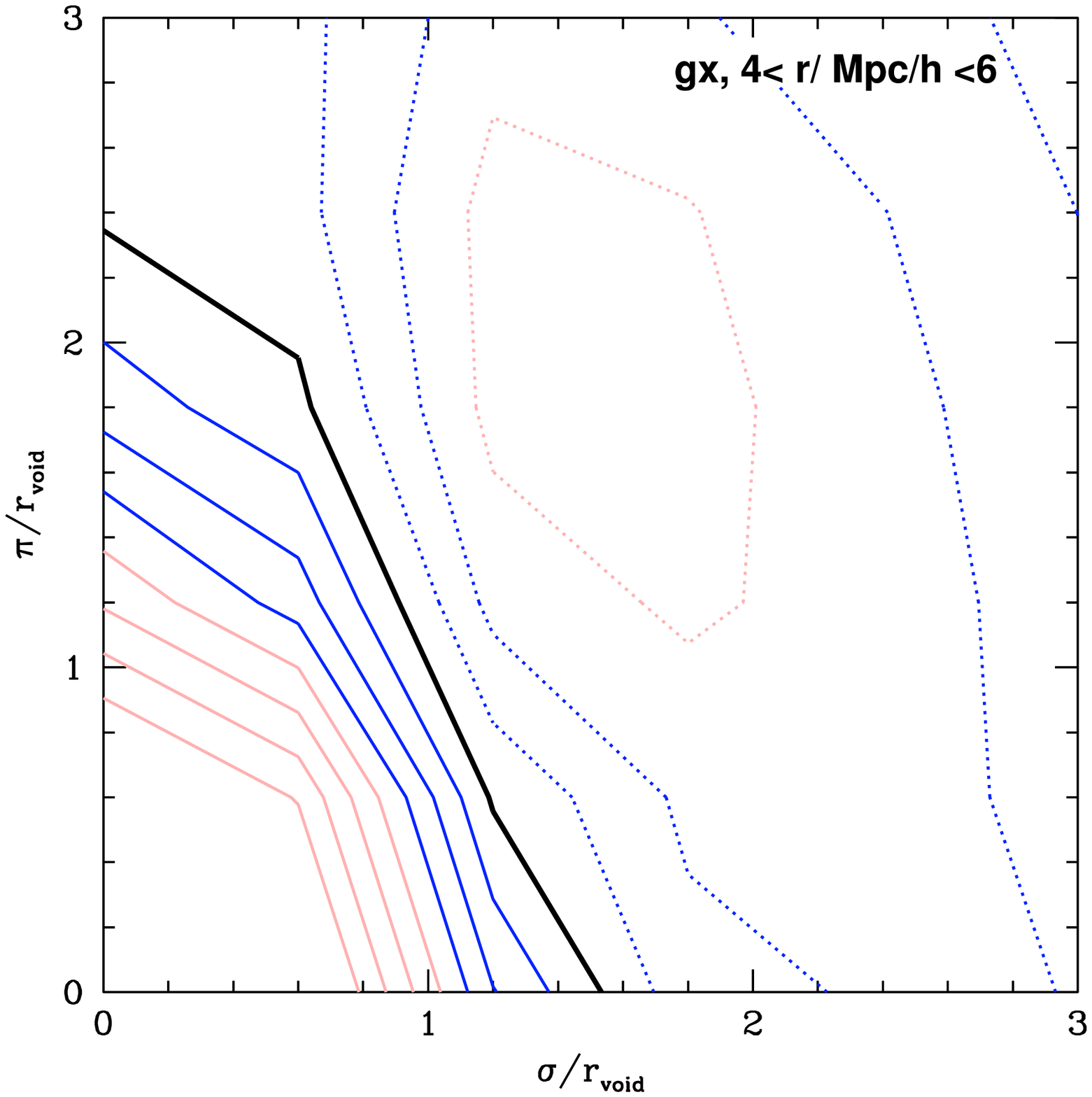,width=7.cm}}
\put(0,145){\psfig{file=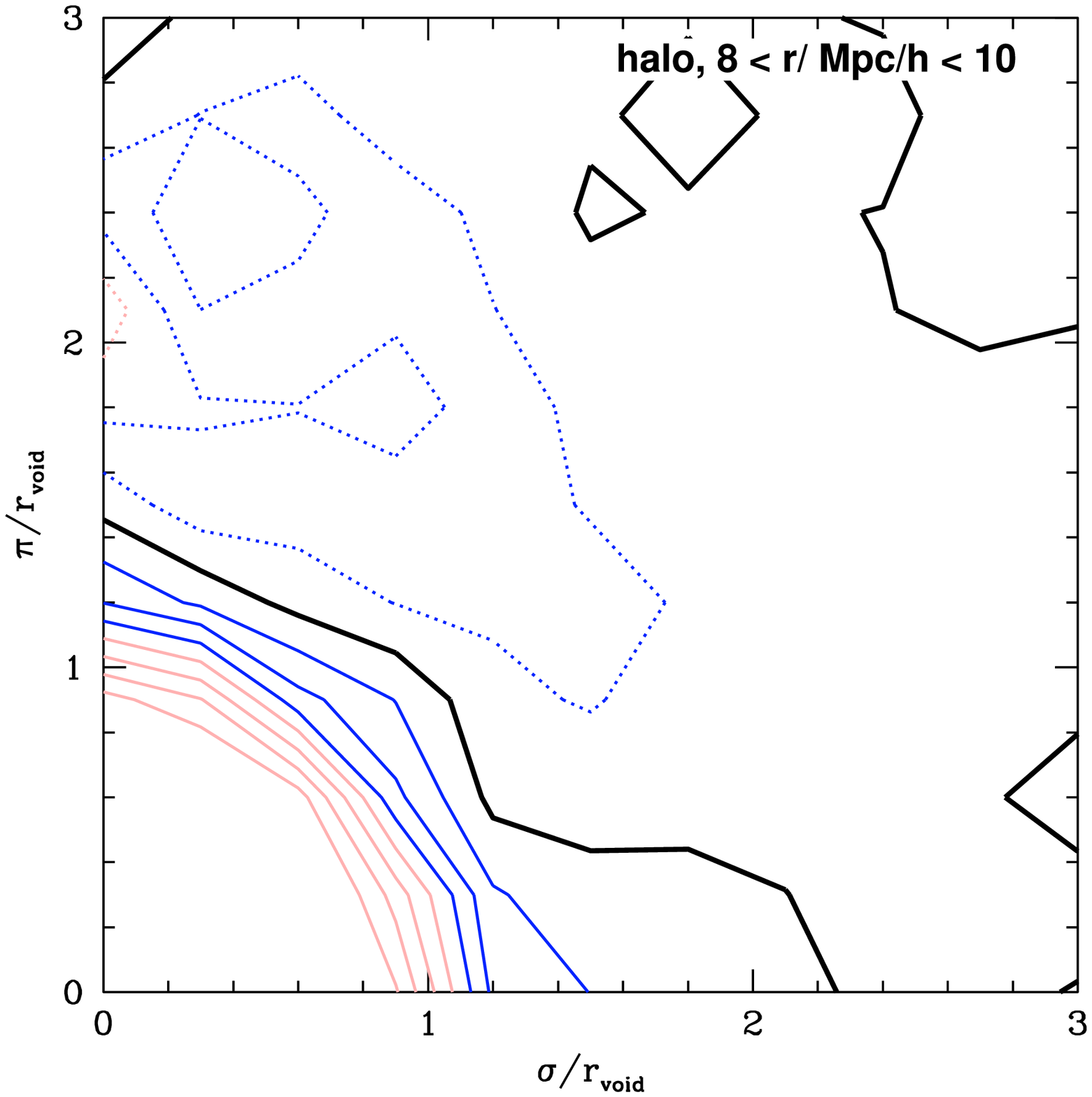,width=7.cm}}
\put(195,145){\psfig{file=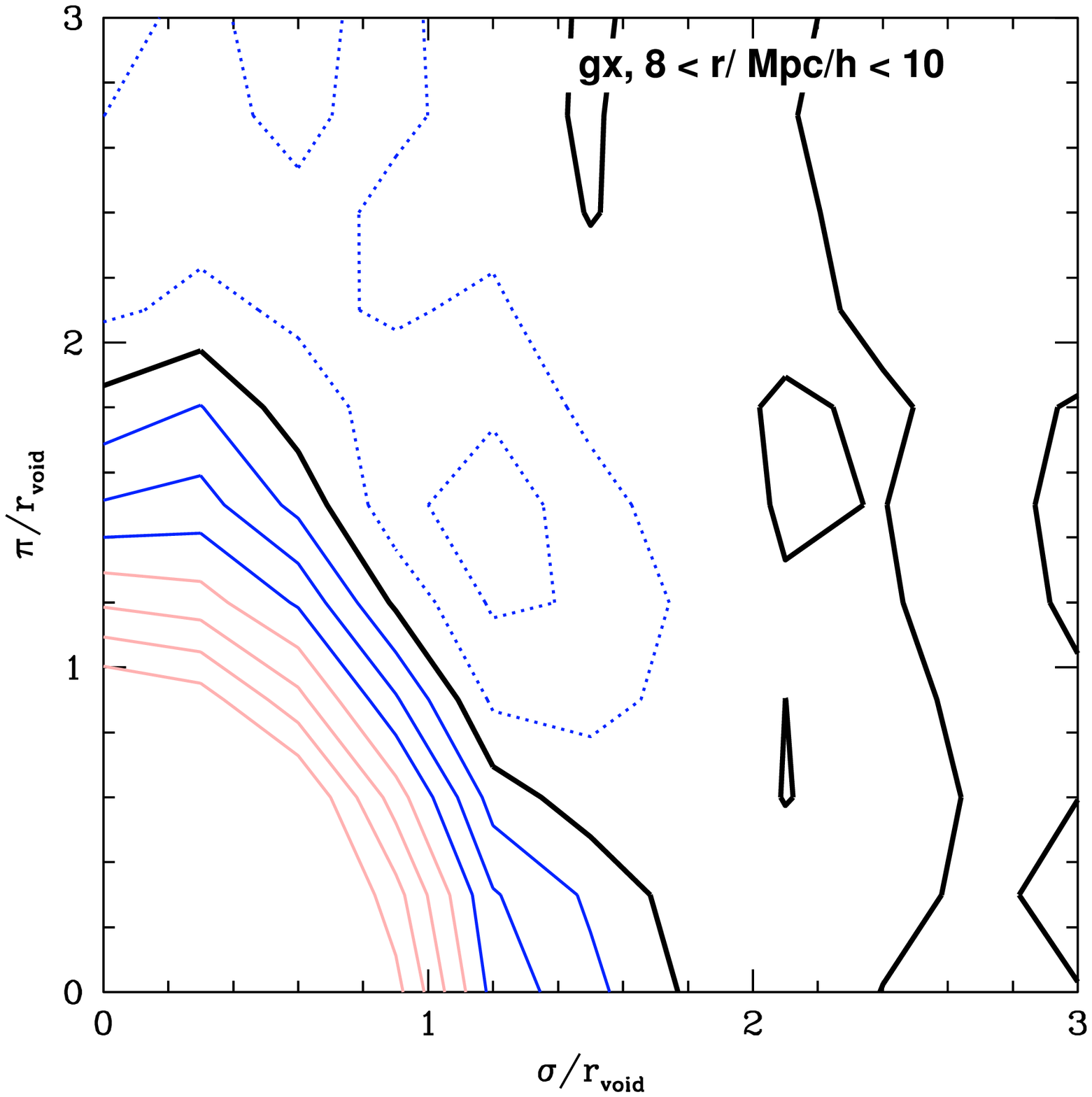,width=7.cm}}
\put(0,340){\psfig{file=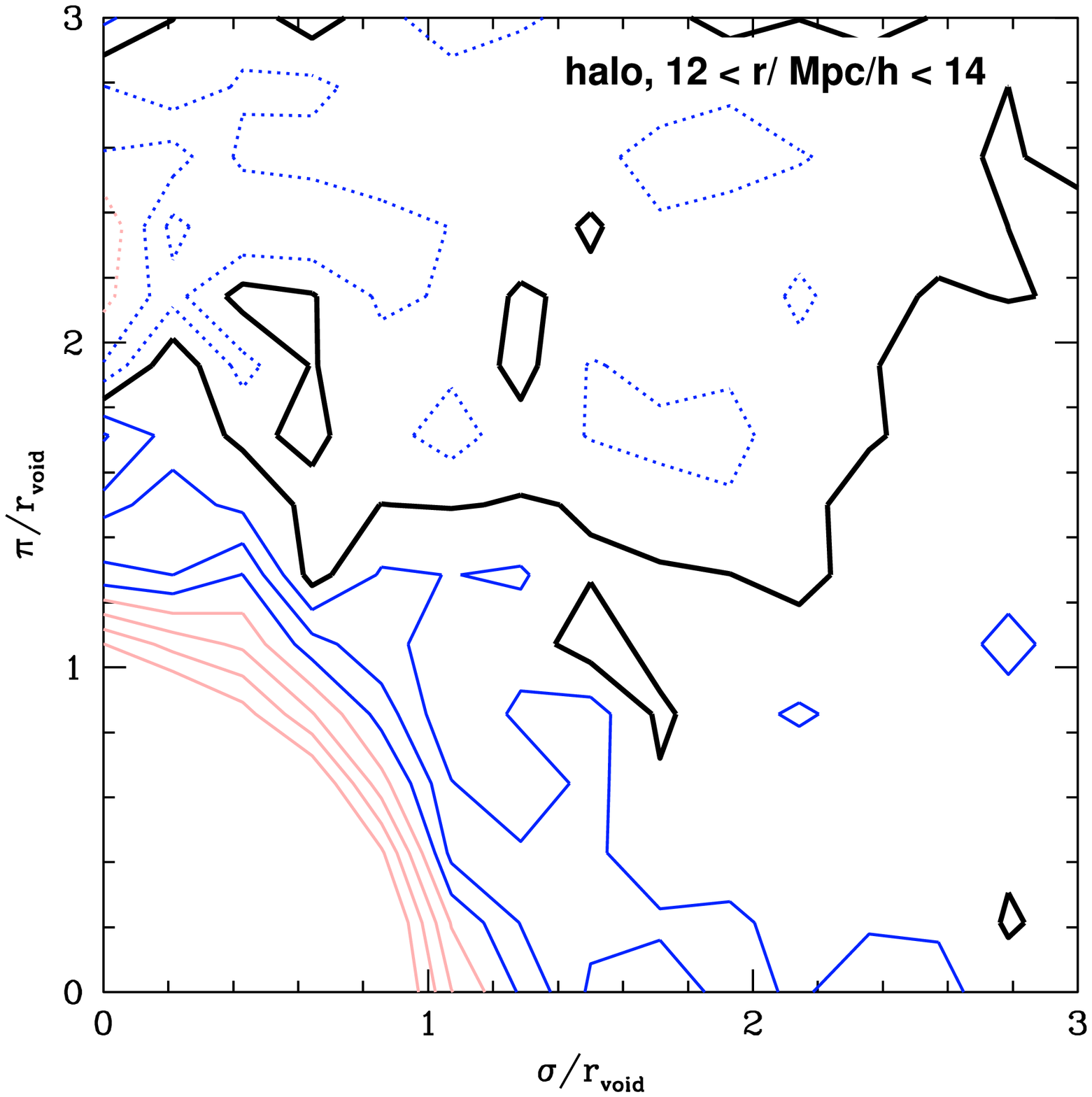,width=7.cm}}
\put(195,340){\psfig{file=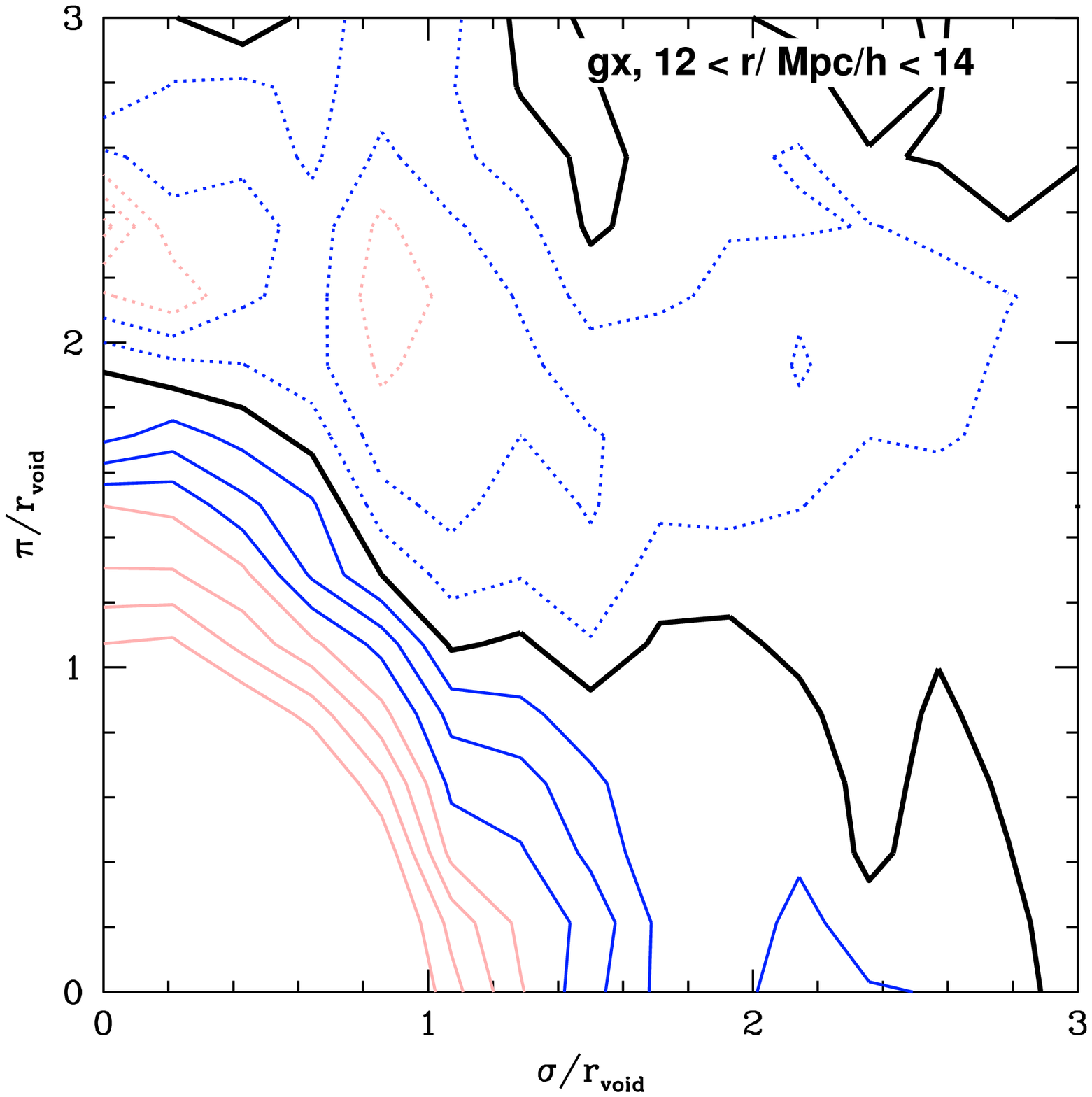,width=7.cm}}
\end{picture}
\caption{
Redshift-space void-halo (left) and void-galaxy (right) cross correlation functions
as a function of the normalised coordinates parallel ($\pi/r_{void}$)
and perpendicular ($\sigma/r_{void}$) to the line of sight
(the z-axis in the numerical simulation).  Here the
centres of voids are those identified in redshift-space.
The solid red (gray) lines represent $\xi(\sigma,\pi)=-0.6,-0.5,-0.4$, solid
blue (black) lines $\xi(\sigma,\pi)=-0.3,-0.2,-0.1$, the thick solid line
corresponds to $\xi(\sigma,\pi)=0.0$.  The dotted blue (black) lines
show $\xi(\sigma,\pi)=0.1,0.2$, and the red (gray) dotted lines,
$\xi(\sigma,\pi)=0.3,0.4,$ and $0.5$.
}
\label{fig:xisp}
\end{figure*}

\begin{table*}
\caption{\small
{
Summary of main differences in the statistics related to voids, when
using galaxies and dark-matter (particles or haloes), in real- and redshift-space.
}}
\begin{tabular}{cll}
\hline
\hline
\noalign{\vglue 0.2em}
Statistic & Real or Redshift-space & Gx., haloes and mass \\
\noalign{\vglue 0.2em}
\hline
\noalign{\vglue 0.2em}
Number density  & Redshift space is higher                   & The number density of Galaxy-defined voids is  \\
                & by up to factor $2.5$ at                   & higher than that of mass-defined voids by as   \\
	        & large void radius. Consistent              & much as two orders of magnitude: consistent at \\
	        & at $r_{\rm void}<4$h$^{-1}$Mpc for mass-   & $r_{\rm void}<4$h$^{-1}$Mpc, but               \\
		& voids.  Halo- and galaxy-defined voids show& no mass-haloes at $r_{\rm void}>12$h$^{-1}$Mpc.\\
		& smaller changes from real- to redshift-    & H$11.5$- and H$12$ haloes in agreement\\
		& space.                                     & with galaxy-defined voids.\\
		& & \\
Void $\xi$	& 				         & Galaxy-defined voids show stronger       \\
		& 				         & auto-correlation functions than      \\
		& 				         & halo-void $\xi$s.      \\
		&				         & Shape of void $\xi$ for voids    \\
		&				         & with different radii are similar,\\
		&				         & and amplitude increases with $r_{\rm void}$.\\
		& & \\
Outflow velocity& Slower outflows and infalls     & Galaxy-defined voids show similar \\
                & in redshift-space (by           & infall and outflow motions   \\
		& about $30$kms$^{-1}$.           & to H$11.5$-defined voids ( differences $<20$kms$^{-1}$)\\
		& & \\
1D $\xi$        & Increase in $\xi$ at $r=r_{void}$ & Galaxy $\xi$ is quite higher  \\
                & is milder when voids and tracers  & than mass $\xi$ outside the void\\
		& are in redshift-space.            & radius.  Inside the void radius both\\
		&                                   & $\xi$s are almost indistinguishable. Good\\
		&                                   & agreement between galaxy- and H$11$-defined voids.\\
		& & \\
2D $\xi$	& Real-space shows no anisotropies& Galaxy-defined voids show larger elongations\\
		& as expected. Redshift-space     & in $\xi(\sigma,\pi)$ than H$11.5$-defined voids.\\
		& elongations extending out to    & Qualitatively, the shapes of contours are\\
		& $\sigma \propto r_{\rm void}$.  & similar for both types of voids.  \\
\noalign{\vglue 0.2em}
\hline
\hline
\end{tabular}\label{table:summary}
\end{table*}

Voids are the opposite of galaxy groups in terms of their surrounding peculiar
velocity field, so we expect to find an elongation of the correlation
function contours along the line of sight. This is expected to affect not only
small separations in the $\sigma$ direction, as is the case with the fingers
of god signatures in the galaxy $\xi(\sigma,\pi)$, 
but also larger $\sigma$ separations.  Figure
\ref{fig:xisp} shows the correlation function contours corresponding
to H$11.5$- and galaxy-defined voids  (left- and right-hand panels
respectively) as a function of normalised separation in the directions
parallel and perpendicular to the line of sight (in our case, the line
of sight coincides with the z-axis in the simulation box).  
The void radius increases from bottom to top panels, 
with values in the ranges shown in the key on each panel.  The
different line types represent different correlation function amplitudes;
solid lines correspond to $\xi\leq0$ and dotted lines to $\xi>0$.  
In order to facilitate the interpretation of this figure, 
and the comparison to results from the auto-correlation function
of galaxies, the reader is referred to figure 4 in Hawkins et al. (2003),
which shows estimates of $\xi(\sigma,\pi)$ for different galaxy types in
the 2dFGRS.

As expected, the void-halo cross-correlations 
show elongated iso-correlation contours
along the line of sight, specially for the $\xi<0.0$ contours.  
This can also be found in the void-galaxy cross correlations,
in particular for the lower correlation function contours, $\xi<-0.3$.
Also, as can be seen in the figure, the contour
patterns contain an unexpected wealth of information which we 
analyse in detail in what follows. 
\begin{itemize}
\item[(i)]
The iso-$\xi$ contours around voids
differ from the elongations seen in
galaxy auto-correlation functions since here the elongated patterns 
extend out to separations $\sigma \simeq r_{void}$, perpendicular 
to the line of sight (ranging from $4$ to $14$h$^{-1}$Mpc from
bottom to top panels). On the other hand, the galaxy auto-correlation
function is significantly distorted only out to $\sigma\simeq 2$h$^{-1}$Mpc (Hawkins
et al. 2003).
\item[(ii)] As expected, Galaxy- and H$11.5$-void $\xi(\sigma,\pi)$ 
show a positive correlation
signal in agreement with the results from the 1-dimensional correlation
functions from the previous section.  The patterns corresponding to positive
correlations show the finger-of-god effect taking place at
the void boundaries. 
\item(iii)
The flattening of the patterns beyond this radius is originated in
random motions taking place in the void walls, which correspond
to a radial velocity dispersion $\sigma_{perp}\simeq 300$kms$^{-1}$ 
(see figure \ref{fig:wallv}).
\item[(iv)] There are usually two $\xi=0$ contours in the 
void $\xi(\sigma,\pi)$
diagrams.  The first one (closer to the origin) corresponds to the increase
in density at the void walls.  
The second $\xi=0$ contour (further away from the origin) is determined by
the end of large scale fluctuations.
\item[(v)]H$11.5$ haloes show an outflow motion of lower amplitude
than that of galaxies.  However, since halos have a larger
velocity dispersion in the void edges, they appear with a 
smaller radial elongation than galaxies.
\end{itemize}

The elongations along the line of sight seen in the 2-dimensional
correlation function of voids can be used to infer the peculiar
velocity field around voids without the need of measuring
peculiar velocities. The latter is a difficult task due to the large
errors involved in the measurement of distances by means independent
of redshift.  We tackle this problem
in a forthcoming paper, Ceccarelli et al. (2005a).

\section{Summary and conclusions}
\label{sec:conc}

In this work we studied the statistical properties of voids 
selected from the distribution of mass, haloes and $B_J<-16$ galaxies in a $\Lambda$CDM 
numerical simulation populated with galaxies
using a GALFORM semi-analytic galaxy formation 
model (Cole et al. 2000). The aim 
of this work was to understand the systematic biases between these 
different void populations (Patiri et al. 2004, Gottl\"ober et al. 2003).  

We find that the number
density of galaxy-defined voids departs from
that of mass-defined voids at $r_{\rm void}\geq 4$h$^{-1}$Mpc. 
However, the fraction of the volume of the simulation box in
galaxy-defined voids and voids identified from haloes with $M>10^{11.5}$h$^{1}$
M$_{\sun}$ (H$11.5$-defined voids) are compatible with one another.
The effect of identifying voids in redshift space increases
the number of voids in the simulation, but this effect is almost
negligible for galaxy- and halo-defined voids.  The fraction of
total volume occupied by galaxy voids in the simulation
reaches the $30\%$ level when considering voids with
$r_{\rm void} >3$h$^{-1}$Mpc.  This fraction is considerably
smaller than results obtained from redshift surveys by
Hoyle \& Vogeley (2002), who find that about $50\%$ of 
the volume in the Universe is in larger voids.  In order to properly assess
whether this disagreement is significant, it must be taken into account
that the volume fraction in voids is very sensitive to the abundance of large
voids in the sample analysed.  Therefore, it would be necessary to 
analyse larger simulations than the one used in this work
to be able to make a more reliable comparison.
Another factor that influences the
volume fraction in voids is the density contrast used to identify
voids; a higher density contrast increases slightly the fraction of
volume.  In addition, the identification of voids in redshift space 
also produces this effect, bringing the results from our simulation even closer
to observational estimates.

The study of the auto-correlation function of voids shows that
voids identified from more massive haloes show a higher
auto-correlation function.  In particular, H$12$-defined voids show
a correlation function which is in agreement with that of
galaxy-defined voids.  Still,
the clustering of voids is very weak, with amplitudes almost always 
lower than $\xi=1$.  We found that the shape of the auto-correlation
function of voids is consistent with being independent of void
radius.  The amplitude, however, is seen to increase with void
size.  By using 100 sets of
random catalogues of voids which satisfy the same exclusion constraints
as galaxy- and halo-defined voids, we demonstrate that the void
auto-correlation function can not be associated solely to an artifact
of the void identification procedure.  The average 
random-voids correlation function
shows a similar shape than that of galaxy-defined voids,
but also shows a lower amplitude 
indicating a possible influence
from hierarchical clustering on the galaxy-defined void auto-correlations.

The study of the outflow velocities around voids 
shows outflow velocities which reach a maximum at a distance,
$d_{\rm vmin} \sim r_{\rm void}$.
At larger separations, $d_{\rm zero}$, the outflow velocities become
moderate infall motions due to the higher density of the void walls and
outflow motions from neighbouring voids.
We find that a very simple relation that describes $d_{\rm vmin}$ and
$d_{\rm zero}$ as a function of void radius, with parameters that can be found
in table 1.  
Table 2 shows the linear fit parameters for the relation between void
radius and maximum outflow motions, $r_{\rm void}$ vs. $v_{\rm min}$.
We searched for differences in the velocity field of galaxies and dark-matter haloes
around voids, and found that galaxy-defined voids show similar 
outflows to H$11.5$-defined voids.
Also, the use of redshift space data lowers the
extreme values of velocities around voids, that is, both outflows
and infalls appear less important than in real-space.
We also analysed the galaxy velocity dispersion at the void walls, and
find a systematically larger velocity dispersion in the direction parallel to
the void walls with respect to the radial direction (a $\simeq 10-20\%$ 
difference).

We find that the 
cross-correlation functions between voids and galaxies, haloes and mass
show negative constant
values out to separations comparable with the void size.  At 
larger separations the galaxy-void - galaxy correlation function level 
increases to approach the auto-correlation function of the galaxies.
The amplitude of the correlation function of galaxies near the
centres of galaxy-defined voids is slightly lower than that of the mass around 
mass-defined voids.

We also studied the distortion pattern observed in
$\xi(\sigma,\pi)$ and found an elongation along the line of sight.  This
elongation differs from that found in the galaxy auto-correlation function, as this
effect extends out to large separations perpendicular to the line of sight.
The study of these elongations would allow us to obtain properties of the
velocity field around voids using only measured redshifts, which are
much easier to obtain observationally than galaxy peculiar velocities.
The study of $\xi(\sigma,\pi)$ not only provides information about
the peculiar velocity field around voids, but also on finger-of-god
motions at the void walls through the elongation of $\xi>0.1$ contours
along the line of sight, at separations $\pi \simeq 2-3 r_{\rm void}$.

We find that the study of voids selected from the distribution
of galaxies can provide biased results which can not easily be reconciled
with results from dark-matter haloes.  For instance, the number density 
of voids as a function of radius for galaxy-defined voids is in agreement
with that of H$12$-defined voids for large void radii, and with H$11.5$ voids
at small radii.  Also, the void auto-correlation function of
galaxy-defined voids is in better agreement with H$12$-defined voids, than with H$11.5$-defined voids;
the velocity field of galaxies is in better agreement with H$11.5$-defined voids.
However, the elongations in the galaxy-void 
$\xi(\sigma,\pi)$ are qualitatively
matched by H$11.5$-haloes.  Therefore care must be taken
when making an interpretation based on galaxy-void data.
We present table \ref{table:summary}
as an overall summary of the differences between properties of voids
selected from galaxies and dark-matter haloes in real- and
redshift-space.

It can be argued that, observationally, the easiest way to obtain the
velocity field surrounding voids is the 2-dimensional
correlation function $\xi(\sigma,\pi)$.  
We analyse whether it is possible to estimate directly the
peculiar velocity field around voids from peculiar velocity data,
and also measure $\xi(\sigma,\pi)$
using large observational datasets such as 2dFGRS and SDSS 
in the next step of this work, which at present is being carried out
in a forthcoming paper (Ceccarelli et al., 2005a), where we pay
special attention to many observational biases such as distance measurement
errors, irregular angular completeness masks and flux limit effects.

\section*{Acknowledgments}
This work was supported in part by the ESO-Chile Joint
Committee,
NDP was supported by a Proyecto Postdoctoral Fondecyt
no. 3040038. DGL and LC are supported by CONICET. We thank
the Durham group for providing the semi-analytic galaxy
formation and simulation dark-matter outputs used
in this work.  We thank the Referee for helpful comments
and suggestions.


\begin{thebibliography}{}
\bibitem[]{5}
Bahcall, N.A., \& Cen, R. 1992, ApJ, 398, 81.	
\bibitem[]{10}
Ceccarelli, L., et al., 2005a, in preparation (C05).
\bibitem[]{20}
Ceccarelli, L., Valotto, C.A., Lambas, D.G., Padilla, N.D., Giovanelli, R., \& Haynes, M., 
2005b, accepted for publication in ApJ.
\bibitem[]{30}
Cole, S., Lacey, C.G., Baugh, C.M., \& Frenk, C.S., 2000, MNRAS, 319, 168.	
\bibitem[]{40}
Croton, D.J., et al. (the 2dFGRS Team), 2004, submitted to MNRAS, Astro-ph/0401406.
\bibitem[]{43}
Einasto, J., Einasto, M., \& Gramman, M., 1989, MNRAS, 238, 155.
\bibitem[]{44}
El-Ad H., \& Piran, T., 1997, ApJ, 491, 421.
\bibitem[]{45}
El-Ad H., \& Piran, T., 2000, MNRAS, 313, 553.
\bibitem[]{46}
Geller, M.J., \& Huchra, J.P., 1989, Sci, 246, 897.	
\bibitem[]{48}
Ghigna, S., Bonometto, S.A., Retzlaff, J., Gottloeber, S., \& Murante, G., 1996, ApJ, 469, 40.	
\bibitem[]{50}
Goldberg, D.M., Jones, T.D., Hoyle, F., Rojas, R.R., Vogeley, M.S., \& Blanton, M.R.,
2004, submitted to ApJ, Astro-ph/0406527.
\bibitem[]{55}
Gottl\"ober S., Lokas, E.L., Klypin, A., \& Hoffman, Y., 2003, MNRAS, 344, 715.
\bibitem[]{57}
Gregory S.A., Thompson, L.A., 1978, ApJ, 222, 784. 
\bibitem[]{70}
Hawkins, E., et al. (the 2dFGRS Team), 2003, MNRAS, 346, 78.	
\bibitem[]{58}
Hoffman, Y., \& Shaham, J., 1982, ApJ, 262, L23.
\bibitem[]{75}
Hoyle, F., Outram, P.J., Shanks, T., Boyle, B.J., Croom, S.M., \& Smith, R.J.,
2002, MNRAS, 332, 311.	
\bibitem[]{59}
Hoyle, F., \& Vogeley, M.S., 2002, ApJ, 566, 641.
\bibitem[]{60}
Hoyle, F., Rojas, R.R., Vogeley, M.S., \& Brinkmann, J.,
2003, submitted to ApJ, Astro-ph/0309728.
\bibitem[]{63}
Joeveer, M., Einasto, J., Tago, E., 1978, MNRAS, 185, 357.	
\bibitem[]{64}
Kirshner, R. P., Oemler, A., Schechter, P. L., Shectman, S. A., 1981, ApJ, 248, 57.	
\bibitem[]{65}
Loveday, J., Maddox, S.J., Efstathiou, G., \& Peterson, B.A., 1995, ApJ, 442, 457.	
\bibitem[]{73}
M\"uller, V., Arbabi-Bidgoli, S., Einasto, J., \& Tucker, D., 2000, MNRAS, 318, 280.	
\bibitem[]{80}
Padilla, N.D., Merch\'an, M., Valotto, C., Lambas, D.G., \& Maia, M., 2001, ApJ.
\bibitem[]{81}
Padilla, N.D. \& Baugh, C.M., 2003, 343, 796.
\bibitem[]{82}
Padilla, N.D. \& Lambas, D.G., 2003a, MNRAS, 342, 532.
\bibitem[]{84}
Padilla, N.D. \& Lambas, D.G., 2003b, MNRAS, 342, 519.
\bibitem[]{86}
Patiri, S.G., Betancort-Rijo, J.E., \& Prada, F., 2004, submitted to ApJ, Astro-ph/0407513.
\bibitem[]{88}
Peebles, P.J.E., 1982, ApJ, 257, 438.
\bibitem[]{89}
Peebles, P.J.E., 2001, ApJ, 557, 495.
\bibitem[]{90}
Percival, W., et al. (the 2dFGRS Team), 2004, MNRAS Accepted, Astro-ph/0406513.
\bibitem[]{95}
Plionis, M., \& Basilakos, S., 2002, MNRAS, 330, 399.
\bibitem[]{100}
Ryden, B.S., 1995, ApJ, 452, 25.
\bibitem[]{110}
Spergel, D.N., et al. (the WMAP team), 2003, ApJS, 148, 175.	
\bibitem[]{120}
van de Weygaert, R., Sheth, R., \& Platen, E., 2004, Proceedings IAU Colloquium, 195.
\bibitem[]{125}
Vogeley, M.S., Geller, M.J., Park, C., \& Huchra, J.P., 1994,  AJ, 108, 745.	
\bibitem[]{130}
White, S. D. M., 1979, MNRAS, 186, 145.	
\end{thebibliography}
\end{document}